\documentclass[12pt,fleqn,a4paper]{article}
\usepackage{a4wide}
\usepackage{epsfig}
\usepackage{slashed}
\usepackage[intlimits]{amsmath}
\usepackage{amssymb}
\usepackage[small]{caption}
\usepackage{cite}
\numberwithin{equation}{section}
\usepackage{color}

\newcommand{\dlogmu}{\mu\frac{d}{d\mu}}
\newcommand{\MS}{$\overline{\text{MS}}$}

\begin{document}

\title{
\vskip-3cm{\baselineskip14pt
\centerline{\normalsize TTP 10-16\hfill}
} 
\vskip1.5cm 
\boldmath
\textbf{$\epsilon_K$ at Next-to-Next-to-Leading Order:\\
  The Charm-Top-Quark Contribution}}
\unboldmath

\author{
  {$\text{Joachim Brod}^{a,b,c}$ and $\text{Martin Gorbahn}^{b,c}$}\\[1em]
  {\normalsize $^{a}$Institut f\"ur Theoretische Teilchenphysik,}\\
  {\normalsize  Universit\"at Karlsruhe, D-76128 Karlsruhe, Germany}\\[1em] 
  {\normalsize $^{b}$Excellence Cluster Universe, 
    Technische Universit\"at M\"unchen,}\\
  {\normalsize Boltzmannstra\ss{}e 2, D-85748 Garching}\\[1em] 
  {\normalsize $^{c}$Institute for Advanced Study, Technische
    Universit\"at M\"unchen, }\\ 
  {\normalsize  Arcisstra\ss{}e 21, D-80333 M\"unchen, Germany}\\[1em]
}

\date{}

\maketitle


\begin{abstract}
  We perform a next-to-next-to-leading order (NNLO) QCD analysis of
  the charm-top-quark contribution $\eta_{ct}$ to the effective
  $|\Delta S|=2$ Hamiltonian in the Standard Model. $\eta_{ct}$
  represents an important part of the short distance contribution to
  the parameter $\epsilon_K$. We calculate the three-loop anomalous
  dimension of the leading operator $\tilde Q_{S2}$, the three-loop
  mixing of the current-current and penguin operators into $\tilde
  Q_{S2}$, and the corresponding two-loop matching conditions at the
  electroweak, the bottom-quark, and the charm-quark scale. As our
  final numerical result we obtain $\eta_{ct} = 0.496 \pm 0.047$,
  which is roughly $7 \%$ larger than the next-to-leading-order (NLO)
  value $\eta_{ct}^{\text{NLO}} = 0.457 \pm 0.073$. This results in a
  prediction for $|\epsilon_K| = (1.90 \pm 0.26)\times 10^{-3}$, which
  corresponds to an enhancement of approximately $3\%$ with respect
  to the value obtained using $\eta_{ct}^{\text{NLO}}$.
\end{abstract}


\section{Introduction}\label{sec:intro}

Indirect $CP$ violation in the neutral Kaon system was discovered by
Christenson, Cronin, Fitch and Turlay in 1964, who observed the decay
of a $K_L$ into two pions \cite{Christenson:1964fg}. This decay would
be forbidden in the case of exact $CP$ symmetry. The parameter
$\epsilon_K$ measures indirect $CP$ violation and is defined by
\begin{equation}
  \label{eq:ekexpdef}
  \epsilon_K = \frac{
    \mathcal{A} \left ( K_L \to (\pi \pi)_{I=0} \right )}{
    \mathcal{A} \left ( K_S \to (\pi \pi)_{I=0} \right )}
\end{equation}
via the ratio of the respective decay amplitudes of a $K_L$ and a
$K_S$ decaying into a two-pion state of isospin zero in such a way
that direct $CP$ violation is absent to a good approximation.

The parameter $\epsilon_K$ is measured with high accuracy: The value
quoted by the Particle Data Group is $\epsilon_K=(2.228\pm
0.011)\times 10^{-3}\times e^{i(43.5\pm
  0.7)^{\circ}}$~\cite{PDG2010}. Whereas until about a decade ago the
numerical value of $\epsilon_K$ was used as an input to determine the
Standard Model parameters, nowadays it plays a central role in
constraining models of new physics: The near diagonality of the
Cabibbo-Kobayashi-Maskawa (CKM) matrix leads to a suppression in the
Standard Model, while $\epsilon_K$ can be predicted very reliably.

For the theoretical prediction it is useful to express $\epsilon_K$ in
terms of $\langle \bar K^0 | \mathcal{H}_{f=3}^{|\Delta S|=2} | K^0
\rangle = 2 M_K M_{12}^*$, the matrix element of the $\Delta S =2$
effective Hamiltonian, and write:
\begin{equation}
  \label{eq:ektheo}
  \epsilon_K = 
  e^{i \phi_\epsilon} \sin \phi_\epsilon 
  \left( \frac {\textrm{Im} (M_{12}^*)}  {\Delta M_K} + \xi \right) \, .
\end{equation}
Here $M_K$ is the neutral Kaon mass and $\Delta M_K$ the Kaon mass
difference, the phase of $\epsilon_K$ is $\phi_\epsilon =
43.5(7)^{\circ}$ \cite{PDG2010} and $\xi = \textrm{Im}
A_0/\textrm{Re} A_0 \simeq 0$ is the imaginary part divided by the
real part of the isospin zero amplitude $A_0 = \mathcal{A} \left ( K_S
  \to (\pi \pi)_{I=0} \right )$. The ratio
$\kappa_\epsilon=|\epsilon_K^{SM}/\epsilon_K(\phi_\epsilon=45^{\circ},
\xi = 0 )|$ encompasses the change of $|\epsilon_K|$ if the values
$\phi_\epsilon = 45^{\circ}$ and $\xi = 0$ are used
in~(\ref{eq:ektheo}), as has been done in most of the older analyses,
instead of the exact values. The authors of
Reference~\cite{Buras:2010pz} give the value of $\kappa_\epsilon =
0.94 \pm 0.02$ in the Standard Model, including in their analysis also
contributions of higher-dimensional operators to the absorptive and
dispersive part of the $K^0$ -- $\bar K^0$ mixing amplitude.

\begin{figure}
  \begin{center}
    \begin{tabular}{cc}
      \includegraphics[scale=0.6]{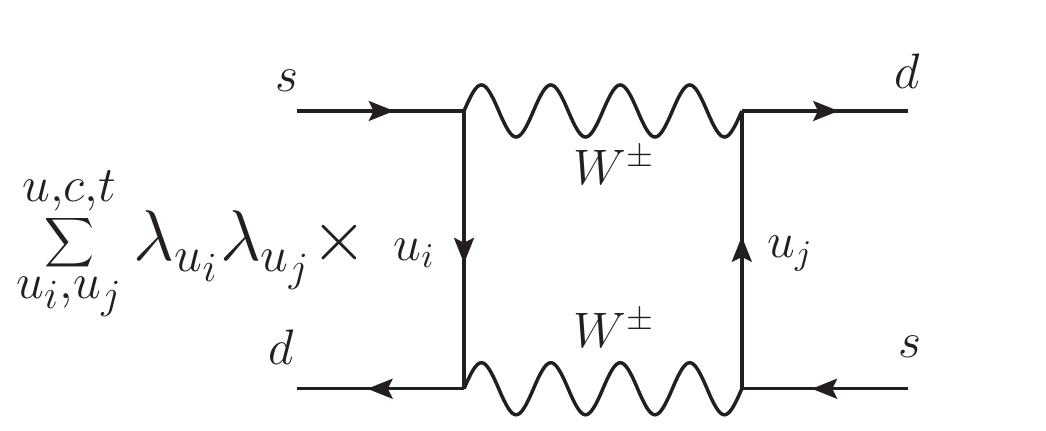} &
      \includegraphics[scale=0.6]{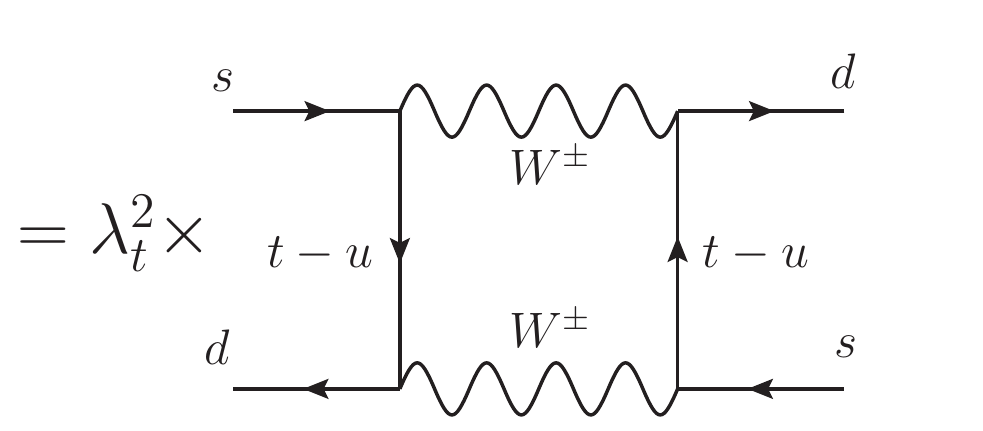} \\
      a) & b) \\
      \includegraphics[scale=0.6]{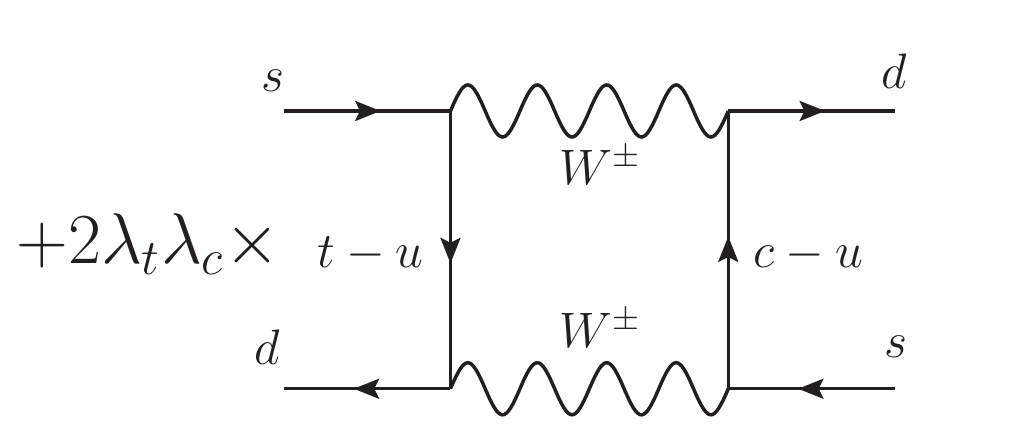} &
      \includegraphics[scale=0.6]{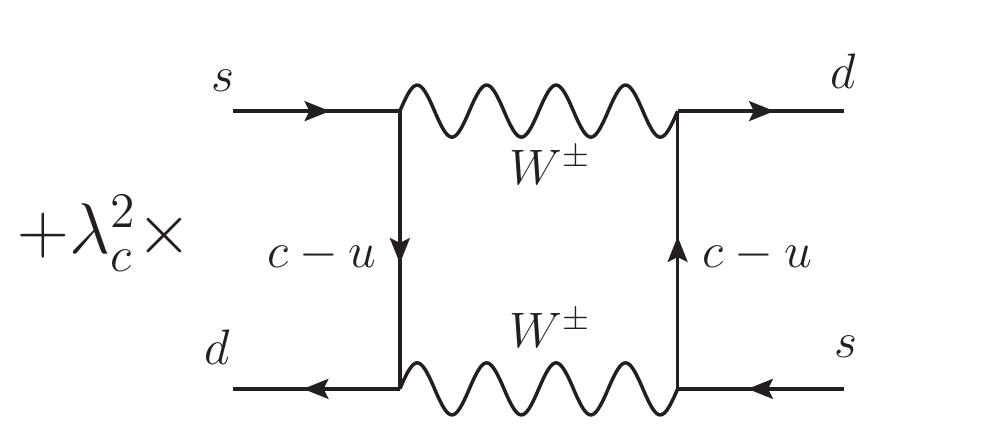} \\
      c) & d)
    \end{tabular}
  \end{center}
  \caption{ The $\Delta S=2$ box-type diagram with internal up,
    charm, and top contributions is expressed as a sum of box-type
    diagrams proportional to $\lambda_t^2$, $\lambda_c^2$, and
    $\lambda_t \lambda_c$, respectively, using the GIM mechanism.}
  \label{fig:leadingbox}
\end{figure}
The box diagram of Fig.~\ref{fig:leadingbox}a gives the leading
contribution to the effective Hamiltonian $\mathcal{H}_{f=3}^{|\Delta
  S|=2}$ and the parameter $M_{12}$. It is proportional to a sum of
loop functions times CKM factors, which, using $\lambda_i = V_{is}^*
V_{id}$ and $x_i = m_i^2/M_W^2$ we can write as:
\begin{equation}
  \label{eq:GIMforS}
  \sum_{u_i,u_j \in \{ u,c,t \}}
  \lambda_{u_i} \lambda_{u_j} 
  \tilde S(x_{u_i},x_{u_j}) =: 
  \lambda_t^2 S(x_t) + 
  \lambda_c^2 S(x_c) + 
  2 \lambda_c \lambda_t S(x_t,x_c) \; ,
\end{equation}
where $\tilde S$ denotes the contributions of the individual box
diagrams. After the Glashow-Iliopoulos-Maiani (GIM) mechanism has been
used to eliminate $\lambda_u = - \lambda_t - \lambda_c$ it comprises
the top-quark contribution -- proportional to $\lambda_t^2$
(Fig.~\ref{fig:leadingbox}b), the charm-quark contribution --
proportional to $\lambda_c^2$ (Fig.~\ref{fig:leadingbox}c), and the
charm-top-quark contribution (Fig.~\ref{fig:leadingbox}d) --
proportional to $\lambda_c \lambda_t$. The resulting loop functions
$S(x_i,x_j) = \tilde S(x_i,x_j) - \tilde S(x_i,0) - \tilde S(0,x_j) +
\tilde S(0,0)$ and $S(x_i) = S(x_i,x_i)$ are suppressed by the
smallness of the quark mass $m_i$ if $x_i$ is significantly smaller
than one. This, together with the severe Cabibbo suppression of the
$CP$ violating top-quark contribution, lets all three contributions
compete in size for $\epsilon_K$:
\begin{equation}\begin{split}
  \label{eq:sizecon}
  &\textrm{Im} \left(
    \lambda_t^2 S(x_t) + 
    \lambda_c^2 S(x_c) + 
    2\lambda_t \lambda_c S(x_t,x_c)
  \right) \\ &\hspace{3cm}\simeq 
  \mathcal{O}\left(\lambda^{10}\right) +
  \mathcal{O}\left(\lambda^{6} \frac{m_c^2}{M_W^2}\right) + 
  \mathcal{O}\left(\lambda^{6} \frac{m_c^2}{M_W^2}
  \log \left(\frac{m_c}{M_W} \right) \right)\; ,
\end{split}\end{equation}
where $\lambda = |V_{us}| \approx 0.2255$. The diagram of
Figure~\ref{fig:leadingbox}a induces a large logarithm $\log m_c/M_W$
only for the charm-top-quark contribution: the large logarithm from
the up quarks in Fig.~\ref{fig:leadingbox}b is power suppressed by
$\Lambda_\textrm{QCD}^2/M_W^2$, while the GIM mechanism cancels a
potential $\log m_c/M_W$ between the diagrams with both one up and one
charm quark and the diagram with only internal charm quarks.

\begin{figure}
  \centering
    \begin{tabular}{cc}
      \includegraphics[scale=0.4]{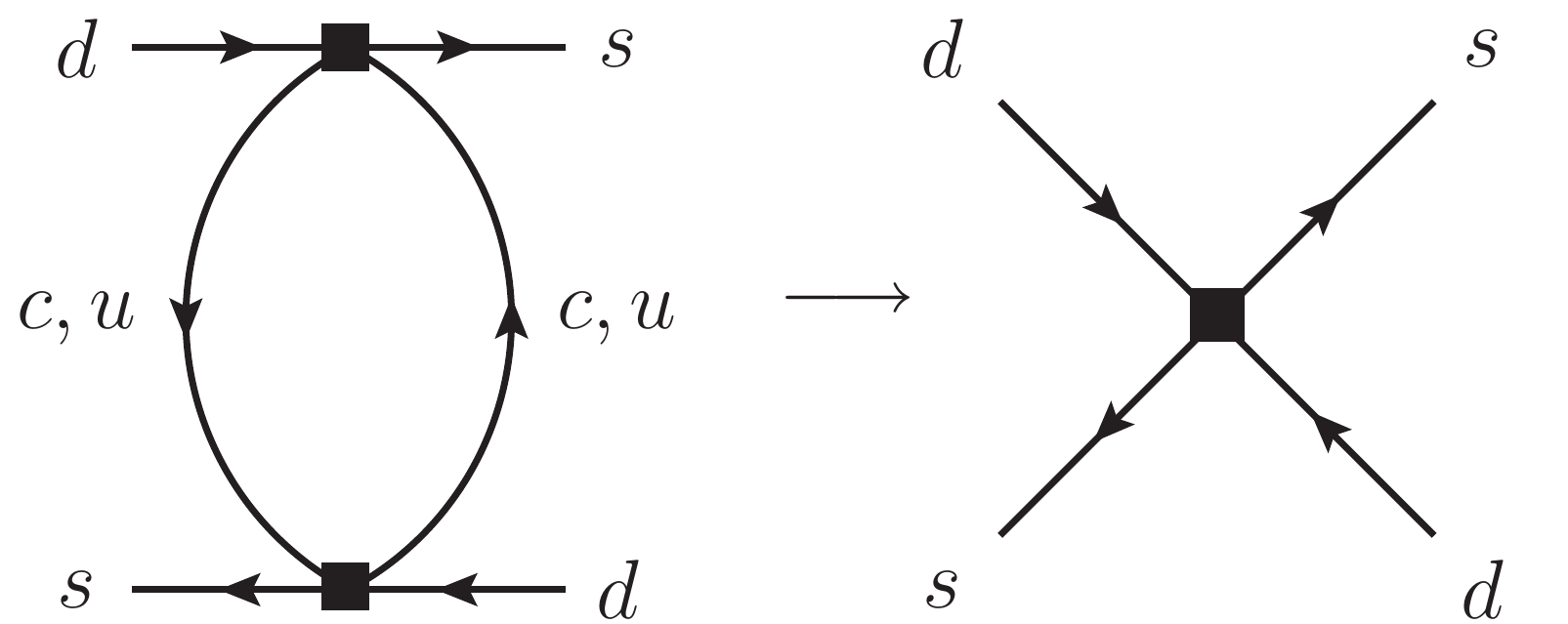} \hspace{1cm} &
      \includegraphics[scale=0.4]{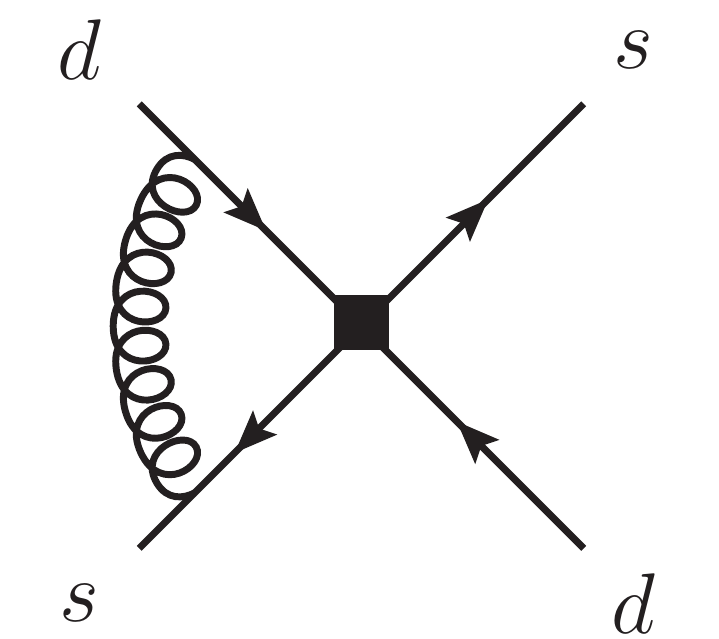} \\
      a) & b) 
    \end{tabular}
    \caption{Dimension-six current-current and penguin operators mix
      at LO into $\tilde Q_{S2}$ with a CKM factor $\lambda_t
      \lambda_c$ in a). Integrating out the charm quark results in
      similar diagrams for the LO and NLO matching of the contribution
      proportional to $\lambda_t \lambda_c$ and $\lambda_c^2$,
      respectively. A sample diagram which is relevant to the LO
      evolution of $\tilde Q_{S2}$ is shown in b).}
  \label{fig:mixinglo}

\end{figure}

This can be reformulated in the language of an effective theory: the
dimension-six penguin as well as the current-current operators, which
have tree-level Wilson coefficients, mix only into the charm-top-quark
contribution, via the bilocal mixing in Fig.~\ref{fig:mixinglo}a, yet
do not induce large logarithms times tree-level Wilson coefficients
proportional to $\lambda_t^2$ and $\lambda_c^2$. QCD corrections do
not change this picture but only induce the well known renormalisation
group effects for the $\Delta S=1$ effective Hamiltonian
\cite{Gorbahn:2004my} and for the $\Delta S=2$ Operator $\tilde
Q_{S2}$ (Fig.~\ref{fig:mixinglo}b). A leading order (LO) analysis of
the charm-quark and top-quark contribution to $\epsilon_k$ then
requires a one-loop calculation both for the matching at $\mu_W$ and
for the running, and for the charm-quark contribution also for the
matching at $\mu_c$ (Fig.~\ref{fig:mixinglo}a). This is in contrast to
the charm-top-quark contribution where a tree-level matching at
$\mu_W$ and $\mu_c$ is sufficient at LO.

After integrating out the charm quark the $\Delta S=2$ effective
Hamiltonian reads
\begin{equation}
    \mathcal{H}^{\Delta S=2}_{f=3} = \frac{G_F^2}{4 \pi^2} M_W^2 \left[
      \lambda_c^2 \eta_{cc} S(x_c) +
      \lambda_t^2 \eta_{tt} S(x_t) +
      2 \lambda_c \lambda_t \eta_{ct} S(x_c,x_t) \right] b(\mu)
    \tilde Q_{S2} + \textrm{H.c.} + \dots
  \label{eq:Hlo}
\end{equation}
where $G_F$ is the Fermi constant and 
\begin{equation}\label{eq:defQS2}
\tilde Q_{S2} = (\overline{s}_L \gamma_{\mu} d_L) \otimes
(\overline{s}_L \gamma^{\mu}d_L) 
\end{equation}
is the leading local four-quark operator that induces the $|\Delta
S|=2$ transition, defined in terms of the left-handed $s$- and
$d$-quark fields. The QCD and logarithmic corrections are known at LO
\cite{Vainshtein:1975xw} and NLO and are parameterised by
$\eta_{cc}=1.43(23)$ \cite{Herrlich:1993yv},
$\eta_{ct}=0.47(4)$\footnote{Our analysis, which uses different inputs
  for the physical parameters and a different error estimate, yields a
  NLO value of $\eta_{ct}=0.457(73)$, see Section~\ref{sec:dnum}.}
\cite{Herrlich:1996vf}, and $\eta_{tt}=0.5765(65)$
\cite{Buras:1990fn}. The parameter $b(\mu)$ is factored out such that
the bag factor
\begin{equation}
  \label{eq:bkpar}
  \hat B_K = \frac{3}{2} b(\mu) 
  \frac{
    \langle \bar K^0 | \tilde Q_{S2} | K^0\rangle}{
    f_K^2 M_K^2} 
\end{equation}
is a renormalisation-group invariant quantity, which can be calculated
on the lattice with high precision -- see for
instance~\cite{Laiho:2009eu}. Here $f_K$ is the Kaon decay constant. 

Finally note that $\mathcal{H}^{\Delta S=2}_{f=3}$ also contains
higher-dimensional operators and current-current operators with up
quarks, as indicated by the ellipses in Equation~(\ref{eq:Hlo}). At LO in
the $1/N_c$ expansion ($N_c$ being the number of colours) only one
higher-dimensional operator is present and its matrix element is
estimated in \cite{Buras:2010pz,Cata:2004ti} to result in a $0.5\%$
enhancement of $\epsilon_K$.

In view of the improvements on the long distance corrections achieved
in the recent years, the short distance contributions should be
reconsidered. In this work we calculate the NNLO corrections to the
charm-top contribution $\eta_{ct}$. The NNLO corrections to the
charm-quark contribution $\eta_{cc}$ will be presented in a
forthcoming publication~\cite{BG}.

This paper is organised as follows. In Section~\ref{sec:Heff} we
define the effective Hamiltonian relevant to $\Delta S=2$
transitions. We present the details of our calculation as well as the
analytic results in Section~\ref{sec:eta3}. The discussion and
numerical evaluation follow in Section~\ref{sec:dnum}. In the appendix
we show how our results transform under a change of the operator
basis.

\section{Effective Hamiltonian for Neutral Kaon Mixing}\label{sec:Heff}

The effective Hamiltonian $\mathcal{H}^{\Delta S=2}_{f=3}$ of
Equation~(\ref{eq:Hlo}) describes the dominant contribution to $\Delta S=2$
processes below the charm-quark mass scale. The loop functions
\begin{align}
  S(x_c)& = x_c + \mathcal{O}(x_c^2)\, ,\label{eq:inamilimbox1}\\
  S(x_t)& = \frac{4x_t-11x_t^2+x_t^3}{4(1-x_t)^2} -
  \frac{3x_t^3\log{x_t}}{2(1-x_t)^3}\, ,\label{eq:inamilimbox2}\\
  S(x_c,x_t)& = - x_c \log x_c + x_c F(x_t) + \mathcal{O}(x_c^2 \log
  x_c)\, ,\label{eq:inamilimbox3}
\end{align}
where the function $F$ is defined as
\begin{equation}
  F(x_t) = \frac{x_t^2-8x_t+4}{4(1-x_t)^2}\log x_t - \frac{3x_t}{4(1-x_t)}
\, ,
\end{equation}
are used as normalisation factors of the three contributions
proportional to $\lambda_c^2$, $\lambda_t^2$, and $\lambda_c
\lambda_t$ in Equation~(\ref{eq:Hlo}). In this normalisation we fix the
charm-quark mass and the top-quark mass to
$m_c=m_c^{\overline{MS}}(m_c)$ and $m_t=m_t^{\overline{MS}}(m_t)$
respectively in $x_c$ and $x_t$. This avoids spurious scale
dependences in $\eta_{ct}$, $\eta_{cc}$, and $\eta_{tt}$, if these
parameters are defined through Equation~\eqref{eq:Hlo}.\footnote{The
  parameters $\eta_{cc}$, $\eta_{tt}$, and $\eta_{ct}$ equal
  $\eta_1^*$, $\eta_2^*$, and $\eta_3^*$, respectively, as defined in
  Reference~\cite{Herrlich:1996vf}.}

\subsection{The Operator Basis}

Above the charm-quark mass scale both the $\Delta S=1$ and $\Delta
S=2$ effective Hamiltonians contribute to the Wilson coefficient of
$\tilde Q_{S2}$ through renormalisation group effects. In the
following we list all operators needed for these effective
Hamiltonians. They can be divided into three classes: Physical
operators, gauge-invariant operators that vanish by the QCD equations
of motion (EOM), and evanescent operators, that vanish algebraically
in four space-time dimensions.

We start with the dimension-six operators, which we choose such that
problems arising from the $\gamma_5$ matrix appearing in closed fermion
loops in the framework of dimensional regularisation do not occur
\cite{Chetyrkin:1997gb}. There are two current-current operators
\begin{equation}\label{eq:Qcurcur}\begin{split}
  Q_1^{qq'}&=(\overline{s}_L\gamma_{\mu}T^a
  q_L)\otimes(\overline{q}'_L\gamma^{\mu}T^a d_L)\, ,\\
  Q_2^{qq'}&=(\overline{s}_L\gamma_{\mu}
  q_L)\otimes(\overline{q}'_L\gamma^{\mu}d_L)\, ,\\
\end{split}\end{equation}
where $q_L = \frac{1}{2}(1-\gamma_5)q$ is the left-handed chiral quark
field, and $q$ and $q'$ are either $u$ or $c$. The colour matrices
$T^a$ are normalised such that $\text{Tr}\, T^a T^b =
\delta^{ab}/2$. We use these operators in the linear combination
\begin{equation}\label{eq:Qpm}
  Q_{\pm}^{qq'} = \frac{1}{2}\left( 1\pm\frac{1}{N_c}\right)
  Q_2^{qq'} \pm Q_1^{qq'} = \frac{1}{2} \left( (\overline{s}_L^{\alpha} \gamma_{\mu}
    q_L^{\alpha} )\otimes(\overline{q}_L^{'\beta} \gamma^{\mu}
    d_L^{\beta}) \pm (\overline{s}_L^{\alpha} \gamma_{\mu}
    q_L^{\beta} )\otimes(\overline{q}_L^{'\beta} \gamma^{\mu}
    d_L^{\alpha})\right) \, ,
\end{equation}
where $\alpha$ and $\beta$ are colour indices, and $N_c$ is the number
of colours. The advantage is that the anomalous dimensions in the
subspace of current-current operators are diagonal in this
basis\footnote{This is true beyond LO only with a suitable choice of
  the evanescent operators, see below.}.

We define the QCD penguin operators as
\begin{equation}\begin{split}\label{eq:Qpin}
  Q_3&=(\overline{s}_L\gamma_{\mu}
  d_L)\otimes\sum\nolimits_q(\overline{q}\gamma^{\mu}q)\, ,\\
  Q_4&=(\overline{s}_L\gamma_{\mu}T^a
  d_L)\otimes\sum\nolimits_q(\overline{q}\gamma^{\mu}T^aq)\, ,\\
  Q_5&=(\overline{s}_L\gamma_{\mu_1\mu_2\mu_3}
  d_L)\otimes\sum\nolimits_q(\overline{q}\gamma^{\mu_1\mu_2\mu_3}q)\, ,\\
  Q_6&=(\overline{s}_L\gamma_{\mu_1\mu_2\mu_3}T^a
  d_L)\otimes\sum\nolimits_q(\overline{q}\gamma^{\mu_1\mu_2\mu_3}T^aq)\, ,
\end{split}\end{equation}
where the sum extends over the light quark fields, and we have
introduced the abbreviations $\gamma_{\mu_1\mu_2\mu_3} =
\gamma_{\mu_1}\gamma_{\mu_2}\gamma_{\mu_3}$, etc. 

\begin{figure}[t]
\centering
\includegraphics[width=\textwidth]{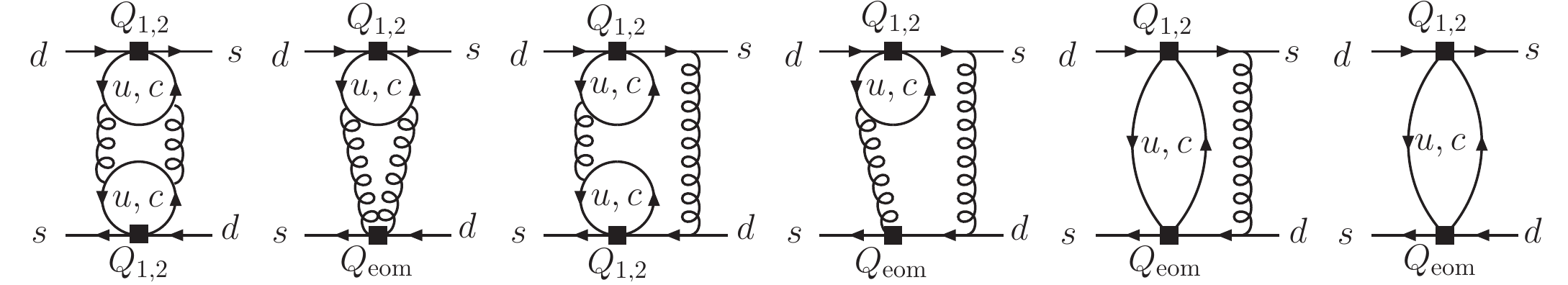}
\caption{ Sample three-loop diagrams with 1PI subdivergences that have
  to be subtracted by insertions of the EOM-vanishing operator. The
  corresponding 1PI one- and two-loop insertions of $Q_{\text{eom}}$ are
  also shown. \label{fig:eom}}
\end{figure}

In order to subtract the divergences of all possible one-particle
irreducible (1PI) subdiagrams of the relevant Green's functions we
need the following gauge-invariant EOM-vanishing operator
\begin{equation}\label{eq:Qeom}
  Q_{\text{eom}}= \frac{1}{g} \bar{s}_L \gamma^{\mu} T^a d_L D^{\nu}
  G^a_{\mu\nu} + Q_4\, ,
\end{equation}
where $D_{\mu}$ denotes the covariant derivative, acting on the gluon
field, and $g^2=4\pi\alpha_s$ is the square of the strong coupling
constant. Sample diagrams are shown in Figure~\ref{fig:eom}.

The operator inducing the effective $|\Delta S|=2$ interactions above
the charm quark scale can be chosen as
\begin{equation}\begin{split}\label{eq:Qtilde7} 
  \tilde{Q}_{7} = \frac{m_c^2}{g^2\mu^{2\epsilon}} (\bar{s}_L^{\alpha} \gamma_{\mu}
  d_L^{\alpha})\otimes (\bar{s}_L^{\beta} \gamma^{\mu} d_L^{\beta})  \, , 
\end{split}\end{equation}
where $\alpha$ and $\beta$ again denote colour indices: Note that,
according to convention, we define the operator with two inverse
powers of the strong coupling constant in order to account for the
logarithm already present at leading order.

The use of dimensional regularisation in a theory involving fermions
implies an infinite-dimensional Dirac algebra. In order to remove all
divergences of the Green's functions, we have to introduce a set of
evanescent operators that are non-zero in $d$ dimensions and vanish
algebraically in four dimensions. At the one-loop level we need
\begin{equation}\begin{split}\label{eq:Eoneloop}
E_1^{qq'(1)}&=(\overline{s}_L\gamma_{\mu_1\mu_2\mu_3}T^a
q_L)\otimes(\overline{q}'_L\gamma^{\mu_1\mu_2\mu_3}T^a d_L) - (16 -
4\epsilon - 4\epsilon^2) Q^{qq'}_1\, ,\\ 
E_2^{qq'(1)}&=(\overline{s}_L\gamma_{\mu_1\mu_2\mu_3}
q_L)\otimes(\overline{q}'_L\gamma^{\mu_1\mu_2\mu_3} d_L) - (16 -
4\epsilon - 4\epsilon^2) Q^{qq'}_2\, ,\\
E_3^{(1)}&=(\overline{s}_L\gamma_{\mu_1\mu_2\mu_3\mu_4\mu_5}
d_L)\otimes\sum\nolimits_q(\overline{q}\gamma^{\mu_1\mu_2\mu_3\mu_4\mu_5}q) +
64Q_3 - 20Q_5\, ,\\
E_4^{(1)}&=(\overline{s}_L\gamma_{\mu_1\mu_2\mu_3\mu_4\mu_5}T^a
d_L)\otimes\sum\nolimits_q(\overline{q}\gamma^{\mu_1\mu_2\mu_3\mu_4\mu_5}T^aq) +
64Q_4 - 20Q_6\, .
\end{split}\end{equation}
At the two-loop level, we use the following four operators: 
\begin{equation}\begin{split}\label{eq:Etwoloop}
E_1^{qq'(2)}&=(\overline{s}_L\gamma_{\mu_1\mu_2\mu_3\mu_4\mu_5}T^a
q_L)\otimes(\overline{q}'_L\gamma^{\mu_1\mu_2\mu_3\mu_4\mu_5}T^a d_L) -
\left ( 256 - 224 \epsilon - \frac{5712}{25} \epsilon^2 \right )
Q^{qq'}_1 \, ,\\  
E_2^{qq'(2)}&=(\overline{s}_L\gamma_{\mu_1\mu_2\mu_3\mu_4\mu_5}
q_L)\otimes(\overline{q}'_L\gamma^{\mu_1\mu_2\mu_3\mu_4\mu_5} d_L) -
\left ( 256 - 224 \epsilon - \frac{10032}{25} \epsilon^2 \right )
Q^{qq'}_2 \, ,\\
E_3^{(2)}&=(\overline{s}_L\gamma_{\mu_1\mu_2\mu_3\mu_4\mu_5\mu_6\mu_7}
d_L)\otimes\sum\nolimits_q(\overline{q}\gamma^{\mu_1\mu_2\mu_3\mu_4\mu_5\mu_6\mu_7}q) +
1280Q_3 - 336Q_5\, ,\\
E_4^{(2)}&=(\overline{s}_L\gamma_{\mu_1\mu_2\mu_3\mu_4\mu_5\mu_6\mu_7}T^a
d_L)\otimes\sum\nolimits_q(\overline{q}\gamma^{\mu_1\mu_2\mu_3\mu_4\mu_5\mu_6\mu_7}T^aq) +
1280Q_4 - 336Q_6\, .
\end{split}\end{equation}
The evanescent operators in the current-current sector are chosen such
that the anomalous dimensions for the operators $Q_{\pm}^{qq'}$ are
diagonal through NNLO~\cite{Buras:2006gb}. The remaining operators are
chosen as in Reference~\cite{Chetyrkin:1997gb}. 

In addition to the operator $\tilde Q_7$ with the colour structure
$(\bar{s}_L^{\alpha} \gamma_{\mu} d_L^{\alpha})\otimes
(\bar{s}_L^{\beta} \gamma^{\mu} d_L^{\beta})$, the dimension-six and
dimension-eight operators will also mix into an operator with the
colour structure \mbox{$(\bar{s}_L^{\alpha} \gamma_{\mu}
  d_L^{\beta})\otimes (\bar{s}_L^{\beta} \gamma^{\mu}
  d_L^{\alpha})$}. In four space-time dimensions, the latter is
related to the former structure by a Fierz transformation. The
difference of these structures is therefore evanescent, and
correspondingly we introduce an evanescent operator of the following
form:
\begin{equation}\label{eq:EtildeF} 
  \tilde{E}_{F} = \frac{m_c^2}{g^2\mu^{2\epsilon}} (\bar{s}_L^{\alpha} \gamma_{\mu}
  d_L^{\beta})\otimes (\bar{s}_L^{\beta} \gamma^{\mu} d_L^{\alpha}) -
  \tilde{Q}_{7} \, . 
\end{equation}
We choose the remaining evanescent dimension-eight operators to be
\begin{equation}\begin{split}
  \tilde{E}_{7}^{(1)} & = \frac{m_c^2}{g^2\mu^{2\epsilon}}
  (\bar{s}_L^{\alpha} \gamma_{\mu_1\mu_2\mu_3}
  d_L^{\alpha})\otimes (\bar{s}_L^{\beta} \gamma^{\mu_1\mu_2\mu_3}
  d_L^{\beta}) - (16 - 4 \epsilon - 4 \epsilon^2) \tilde{Q}_{7} \, , \\   
  \tilde{E}_{8}^{(1)} & = \frac{m_c^2}{g^2\mu^{2\epsilon}}
  (\bar{s}_L^{\alpha} \gamma_{\mu_1\mu_2\mu_3}
  d_L^{\beta})\otimes (\bar{s}_L^{\beta} \gamma^{\mu_1\mu_2\mu_3}
  d_L^{\alpha}) - (16 - 4 \epsilon - 4 \epsilon^2) (\tilde{Q}_{7} + \tilde{E}_F) \, , \notag
\end{split}\end{equation}
\begin{equation}\begin{split}
  \tilde{E}_{7}^{(2)} & = \frac{m_c^2}{g^2\mu^{2\epsilon}}
  (\bar{s}_L^{\alpha} \gamma_{\mu_1\mu_2\mu_3\mu_4\mu_5}
  d_L^{\alpha})\otimes (\bar{s}_L^{\beta} \gamma^{\mu_1\mu_2\mu_3\mu_4\mu_5}
  d_L^{\beta}) - (256 - 224 \epsilon - \frac{108\, 816}{325} \epsilon^2) \tilde{Q}_{7}\, , \\ 
  \tilde{E}_{8}^{(2)} & = \frac{m_c^2}{g^2\mu^{2\epsilon}}
  (\bar{s}_L^{\alpha} \gamma_{\mu_1\mu_2\mu_3\mu_4\mu_5}
  d_L^{\beta})\otimes (\bar{s}_L^{\beta} \gamma^{\mu_1\mu_2\mu_3\mu_4\mu_5}
  d_L^{\alpha}) - (256 - 224 \epsilon - \frac{108\, 816}{325}
  \epsilon^2) (\tilde{Q}_{7} + \tilde{E}_F) \, , \\
  \tilde{E}_{7}^{(3)} & = \frac{m_c^2}{g^2\mu^{2\epsilon}}
  (\bar{s}_L^{\alpha} \gamma_{\mu_1\mu_2\mu_3\mu_4\mu_5\mu_6\mu_7}
  d_L^{\alpha})\otimes (\bar{s}_L^{\beta} \gamma^{\mu_1\mu_2\mu_3\mu_4\mu_5\mu_6\mu_7}
  d_L^{\beta}) - 4096 \tilde{Q}_{7} \, , \\   
  \tilde{E}_{8}^{(3)} & = \frac{m_c^2}{g^2\mu^{2\epsilon}}
  (\bar{s}_L^{\alpha} \gamma_{\mu_1\mu_2\mu_3\mu_4\mu_5\mu_6\mu_7}
  d_L^{\beta})\otimes (\bar{s}_L^{\beta} \gamma^{\mu_1\mu_2\mu_3\mu_4\mu_5\mu_6\mu_7}
  d_L^{\alpha}) - 4096 (\tilde{Q}_{7} + \tilde{E}_F) \, . 
\end{split}\end{equation}
This choice ensures that $\tilde Q_{S2}$ will have the same anomalous
dimension as $Q_+$ up to NNLO. It is given explicitly here for the
first time. 

\subsection{Effective Hamiltonian}

We obtain the effective Hamiltonian valid between the electroweak and
the bottom-quark scale by removing the top quark and the $W$ boson as
dynamical degrees of freedom from the Standard Model. It reads in
terms of the renormalised Wilson coefficients
\begin{align}\label{eq:lags2}
  \mathcal{H}_{f=5}^{\text{eff}} &= \frac{4 G_F}{\sqrt{2}}
  \sum_{i=+,-,3}^6 C_i \left[ \sum_{j=+,-} Z_{ij} \sum_{k,l =u,c}
    V_{ks}^\ast V_{ld} Q_j^{kl} -
    \lambda_t \sum_{j=3}^6 Z_{ij} Q_j  \right] \notag \\
  & + \frac{G_F^2}{4\pi^2} \lambda_t^2 \tilde{C}_{\text{S2}}^{t}
  \tilde{Z}_{\text{S2}} \tilde{Q}_{\text{S2}} + 8 G_F^2 \lambda_c
  \lambda_t \left[ \sum_{k=+,-} \sum_{l=+,-,3}^{6} C_k C_l
    \hat{Z}_{kl,7} + \tilde{C}_7 \tilde{Z}_{77} \right]
  \tilde{Q}_{7} + \text{h.c.}\, .
\end{align}
Here the first line represents the $|\Delta S|=1$ part of the
effective Hamiltonian, whereas the second line contains the $|\Delta
S|=2$ contributions. The first term in the second line is related to a
single insertion of $\tilde{Q}_{\text{S2}}$, induced by the top quark
contribution to the Standard Model amplitude. The remaining terms
arise from the mixing of insertions of two $|\Delta S|=1$ operators
into the operator $\tilde{Q}_{7}$. The GIM mechanism leads to the
absence of a $\lambda_c^2$ contribution to the Wilson coefficient of
$\tilde{Q}_{7}$. The renormalisation constants $Z$ are defined such
that any renormalised effective amplitude, of the form
$\mathcal{A}_{\text{eff}}=C_i(\mu)Z_{ij} \langle Z Q_j \rangle_R + (
  C_{k}C_{k'}\hat{Z}_{kk',l} + \tilde{C}_{k}\tilde{Z}_{kl} )
  \langle Z \tilde{Q}_{l} \rangle_R$,
is finite and implicitly includes the contribution of evanescent
operators. Here angle-brackets denote matrix elements between initial
and final states $i$ and $f$, respectively, i.e. $\langle Q_j\rangle =
\langle f| Q_j |i\rangle$. $Z$ denotes the wave function
renormalisation of the fields in the operator, so that $\langle
ZQ_i\rangle_R$ are the renormalised matrix elements of the bare
operator $Q_i^{\text{bare}}$, where masses and gauge couplings are
renormalised in the usual way.

The effective Hamiltonian $\mathcal{H}_{f=4}^{\text{eff}}$
valid between the bottom- and the charm-quark scale looks exactly the
same as $\mathcal{H}_{f=5}^{\text{eff}}$. The only difference
is induced by the presence of penguin operators, which explicitly
depend on all light quark fields.

Below the charm-quark scale, the charm quark is removed as a dynamical
degree of freedom. As a consequence, the $|\Delta S|=1$ operators can
now be dropped from the effective Lagrangian, because the matrix
elements of double insertions of these operators are suppressed by
factors of $m_s^2/M_W^2$. The effective Hamiltonian is thus given by
\begin{equation}\label{eq:lagmuc}
  \mathcal{H}_{f=3}^{|\Delta S|=2} = \frac{G_F^2}{4\pi^2} \left[
    \lambda_c^2 \tilde{C}_{S2}^{c}(\mu) + \lambda_t^2
    \tilde{C}_{S2}^{t}(\mu) + \lambda_c \lambda_t \tilde{C}_{S2}^{ct}(\mu)
  \right] \tilde{Z}_{S2} \tilde{Q}_{S2} \, 
\end{equation}
and now only contains the $|\Delta S|=2$ operator $\tilde{Q}_{S2}$
defined in Equation~\eqref{eq:defQS2}. 

\boldmath
\section{Calculation of $\eta_{ct}$}\label{sec:eta3}
\unboldmath

In this section we present the details of the calculation of
$\eta_{ct}$ in the NNLO approximation. We start with the determination
of the initial conditions for the Wilson coefficients at the
electroweak scale.  Afterwards we use the renormalisation group
equations to evolve them down to the charm-quark scale, including the
threshold corrections at the bottom-quark scale. Finally we determine
the charm-top contribution to $\tilde{C}_{S2}^{ct}$ by a matching
calculation at the charm-quark scale.
 
\subsection{Initial Conditions at the Electroweak Scale}

The initial conditions for the Wilson coefficients of the
dimension-six operators are available in the literature. In our basis,
where we can use a naive anticommuting $\gamma_5$, the results up to
second order in $\alpha_s^{(f=5)}$ read\footnote{Here and in the
  following, by the superscript in brackets we display explicitly the
  number of light quark flavours for which $\alpha_s$ is defined. }
\begin{equation}\begin{split}\notag
    C_\pm (\mu) & = 1 \pm \frac{1}{2} \left ( 1 \mp \frac{1}{3}
    \right) \left ( 11 + 6 \, L_W \right ) \frac{\alpha_s^{(5)} (\mu)}{4
      \pi} + \Bigg( \frac{1}{18} \left ( 7 \pm 51 \right ) \pi^2 \mp
    \frac{1}{2} \left ( 1
      \mp \frac{1}{3} \right ) T (x_t) \\
    & \quad -\frac{1}{3600} \left ( 135677 \mp 124095 \right )
    -\frac{5}{36} \left ( 11 \mp 249 \right ) L_W + \frac{1}{6} \left
      ( 7 \pm 51 \right ) L_W^2 \Bigg) \left(
      \frac{\alpha_s^{(5)} (\mu)}{4 \pi} \right )^2 \, ,\\
    C_3 (\mu) & = \left( \frac{\alpha_s^{(5)} (\mu)}{4 \pi} \right )^2
    \left(G^t_1(x_t) - \frac{680}{243} - \frac{20}{81}\pi^2 -
      \frac{68}{81} L_W - \frac{20}{27} L_W^2
    \right) \, , \\
    C_4 (\mu) & = \frac{\alpha_s^{(5)} (\mu)}{4 \pi} \left( E^t_0(x_t) -
      \frac{7}{9} + \frac{2}{3} L_W \right) \\
    & \quad  + \left( \frac{\alpha_s^{(5)}
        (\mu)}{4 \pi} \right )^2 \left( E^t_1(x_t) + \frac{842}{243} +
      \frac{10}{81}\pi^2 +
      \frac{124}{27} L_W + \frac{10}{27} L_W^2 \right)\, , \\
\end{split}\end{equation}
\begin{equation}\begin{split}
    C_5 (\mu) & = \left( \frac{\alpha_s^{(5)} (\mu)}{4 \pi} \right )^2 \left
      ( \frac{2}{15} E^t_0 (x_t) - \frac{1}{10} G^t_1 (x_t) +
      \frac{68}{243} + \frac{2}{81}\pi^2 + \frac{14}{81} L_W +
      \frac{2}{27} L_W^2 \right ) \, , \\
    C_6 (\mu) & = \left( \frac{\alpha_s^{(5)} (\mu)}{4 \pi} \right )^2 \left
      ( \frac{1}{4} E^t_0 (x_t) - \frac{3}{16} G^t_1 (x_t) +
      \frac{85}{162} + \frac{5}{108}\pi^2 + \frac{35}{108} L_W +
      \frac{5}{36} L_W^2 \right ) \, .
\end{split}\end{equation}
We have taken the initial conditions for $C_\pm$ from
Reference~\cite{Buras:2006gb}. The initial conditions for $C_3\ldots C_6$
can be found in Reference~\cite{Bobeth:1999mk}, where also the loop
functions $T(x_t)$, $G^t_1(x_t)$, $E^t_0(x_t)$ and $E^t_1(x_t)$ are
defined. Note that in our renormalisation scheme we had to include an
additional finite contribution for $C_4$, as described in
the appendix. We have introduced the abbreviation $L_W = \log
(\mu^2 / M_W^2)$.

\begin{figure}[t]
\centering
\includegraphics[width=10cm]{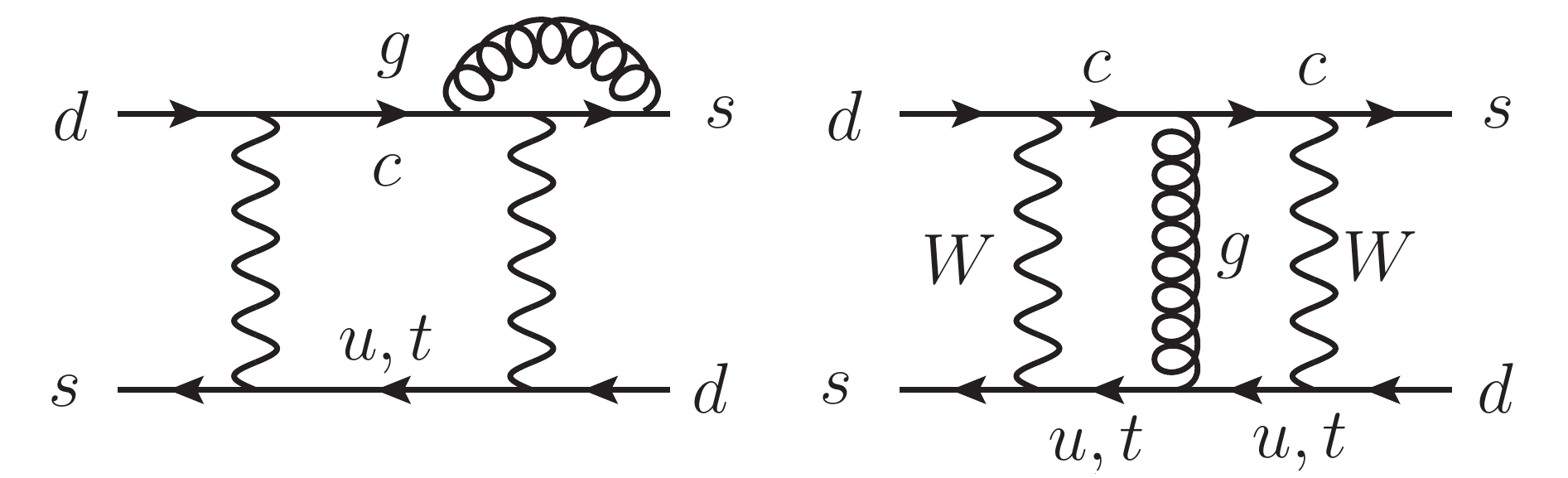}
\caption{ Sample two-loop Feynman diagrams contributing to the matching
  at the electroweak scale. \label{fig:kk2l}}
\end{figure}

With these ingredients, we can now calculate the initial conditions
for the Wilson coefficients of the dimension-eight operators. In order
to match the Green's functions in the Standard Model and the effective
five-flavour theory, we have to compute the finite parts of Feynman
diagrams of the type shown in Figures~\ref{fig:leadingbox}
and~\ref{fig:kk2l}. To this end, we perform a Taylor expansion in the
charm-quark mass of all propagators corresponding to a charm-quark
field. The constant terms cancel because of the GIM mechanism, whereas
the terms proportional to $m_c^2$ give the leading non-vanishing
contribution we are interested in. This procedure leads to massless
vacuum integrals in the effective theory, such that only terms
proportional to tree-level matrix elements remain.  Some of these
terms multiply divergent renormalisation constants and correspond to
infrared divergences in the effective theory. They exactly cancel the
corresponding infrared divergent terms in the Standard Model, leaving
us with a finite result. 

Expanding the dimension-eight Wilson coefficient as
\begin{equation}
  \tilde{C}_7(\mu) = \tilde{C}_7^{(0)}(\mu) +
  \frac{\alpha_s^{(5)}(\mu)}{4\pi} \tilde{C}_7^{(1)}(\mu) +
  \left ( \frac{\alpha_s^{(5)}(\mu)}{4\pi} \right )^2 \tilde{C}_7^{(2)}(\mu)\, ,   
\end{equation}
we obtain the following result:
\begin{equation}\label{eq:C7initLO}\begin{split}
  \tilde{C}_7^{(0)}(\mu) &= 0 \, , \qquad \tilde{C}_7^{(1)}(\mu)
  = F(x_t) + \frac{1}{2} - L_W \, , \\
  \tilde{C}_7^{(2)}(\mu) &= +\frac{5 x_t^3-21 x_t^2+60
      x_t-20}{2 (x_t-1)^3} \log (x_t) L_W\\
  &\quad +\frac{12 x_t^5-34 x_t^4-9 x_t^3-33 x_t^2-116
      x_t+36}{12 (x_t-1)^3} \log ^2(x_t)\\
  &\quad +\frac{-12 x_t^5+27 x_t^4+23 x_t^3+150 x_t^2-108
      x_t+16}{6 (x_t-1)^3 x_t} \log (x_t)\\
  &\quad +\frac{-7800 x_t^4-126499 x_t^3+191248 x_t^2-129749
    x_t+10400}{3900
    (x_t-1)^2 x_t}\\
  &\quad +\frac{6 x_t^6-11 x_t^5-8 x_t^4-29 x_t^3+23 x_t^2-16
      x_t+8}{3 (x_t-1)^2 x_t^2} \text{Li}_2(1-x_t)\\
  &\quad +\frac{6 x_t^4+x_t^3-59 x_t^2-8}{3 x_t^2}
  \zeta_2 -\frac{47 x_t^2-31 x_t+56}{6(x_t-1)^2} L_W -7 L_W^2 
\end{split}\end{equation}
The first line in Equation~\eqref{eq:C7initLO} agrees with the result
obtained already by Herrlich and Nierste in~\cite{Herrlich:1996vf}
after the corresponding change of the renormalisation scheme. The
two-loop result is new.

\subsection{Structure of the Renormalisation Group Equations}\label{sec:rgeeK}

After the determination of the initial conditions for the Wilson
coefficients, the next step is the renormalisation group evolution to
lower scales. The renormalisation group equation relevant for the
Wilson coefficient $\tilde{C}_7$ is given by:
\begin{equation}\label{eq:rgeC7}
\dlogmu \tilde{C}_{7}(\mu) = \tilde{C}_{7}(\mu) \tilde{\gamma}_{77} 
+ \sum_{k=+,-} \sum_{n=+,-,3}^{6} C_{k}(\mu)C_{n}(\mu)\hat\gamma_{kn,7}\, ,
\end{equation}
where $\tilde{\gamma}_{77}$ denotes the anomalous dimension matrix of
the operator $\tilde Q_7$, and $\hat\gamma_{kn,7}$ is the anomalous
dimension tensor, describing the mixing of the dimension-six operators
into $\tilde Q_7$. The matrix $\tilde{\gamma}_{77}$ is decomposed as
$\tilde{\gamma}_{77} = \tilde\gamma_{S2} + 2 \gamma_{m} - 2 \beta$,
where the anomalous dimension of the quark mass $\gamma_{m}$ and the
$\beta$ function are related to the factor $m_c^2/g^2$ in the
definition of the operator $\tilde Q_{7}$. The anomalous dimension
matrix $\tilde\gamma_{S2}$ is defined in terms of the renormalisation
constants $\tilde Z_{S2}$ as
\begin{equation}\label{eq:gammatdef}
\tilde\gamma_{S2} = \tilde Z_{S2}\dlogmu \tilde Z_{S2}^{-1}\, . 
\end{equation}
The explicit expressions for the anomalous dimension matrix in terms
of the renormalisation constants $\tilde Z_{S2}$ are given up to NNLO
by
\begin{equation}\begin{split}
    \tilde\gamma_{S2}^{(0)}&= 2\tilde Z_{S2}^{(1,1)} \, , \qquad
    \tilde\gamma_{S2}^{(1)} = 4\tilde Z_{S2}^{(2,1)} - 2\tilde
    Z_{S2}^{(1,1)}\tilde Z_{S2}^{(1,0)} \, ,\\[2mm]
    \tilde\gamma_{S2}^{(2)}&= 6\tilde Z_{S2}^{(3,1)} - 4\tilde
    Z_{S2}^{(2,1)}\tilde Z_{S2}^{(1,0)} - 2 \tilde
    Z_{S2}^{(1,1)}\tilde Z_{S2}^{(2,0)} \, ,
\end{split}\end{equation}
where we only kept the non-vanishing physical contributions. Here the
superscript $(n,m)$ denotes the $1/\epsilon^m$-pole part of the
$n$-loop contribution. The anomalous dimension tensor is defined as
\cite{Herrlich:1994kh}
\begin{equation}\label{eq:adtdef}
\hat \gamma_{kn,l} = -(\gamma_{kk'}\delta_{nn'}+ \gamma_{nn'}\delta_{kk'})
\hat{Z}_{k'n',l'}\tilde{Z}^{-1}_{l'l}
-\left(\mu\frac{d}{d\mu}\hat{Z}_{kn,l'}\right) \tilde{Z}^{-1}_{l'l}\, .
\end{equation}
The non-vanishing contributions to the physical part of the anomalous
dimension tensor are given in terms of the renormalisation constants
by
\begin{equation}\begin{split}
  \hat\gamma_{kn,l}^{(0)}&=2\hat{Z}_{kn,l}^{(1,1)}\, ,\\[2mm]
  \hat\gamma_{kn,l}^{(1)}&=4\hat{Z}_{kn,l}^{(2,1)} -
  2\hat{Z}_{kn,l'}^{(1,1)}\tilde{Z}_{l'l}^{(1,0)} -
  2\left\{Z_{kk'}^{(1,1)}\delta_{nn'} +
    \delta_{kk'}Z_{nn'}^{(1,1)}\right\}\hat{Z}_{k'n',l}^{(1,0)}\, ,\\[2mm]
  \hat\gamma_{kn,l}^{(2)}&=6\hat{Z}_{kn,l}^{(3,1)}
  - 4\hat{Z}_{kn,l'}^{(2,1)} \tilde{Z}_{l'l}^{(1,0)}
  - 2\hat{Z}_{kn,l'}^{(1,1)} \tilde{Z}_{l'l}^{(2,0)} \\[2mm]
  &\quad - 2\left\{Z_{kk'}^{(1,1)} \delta_{nn'} + \delta_{kk'}
    Z_{nn'}^{(1,1)} \right\} \hat{Z}_{k'n',l}^{(2,0)} -
  4\left\{Z_{kk'}^{(2,1)} \delta_{nn'} + \delta_{kk'} 
    Z_{nn'}^{(2,1)} \right\} \hat{Z}_{k'n',l}^{(1,0)}\, ,
\end{split}\end{equation} 
where the indices $k$, $n$ and $l$ correspond to physical operators only.

\begin{figure}[t]
\centering
\includegraphics[width=\textwidth]{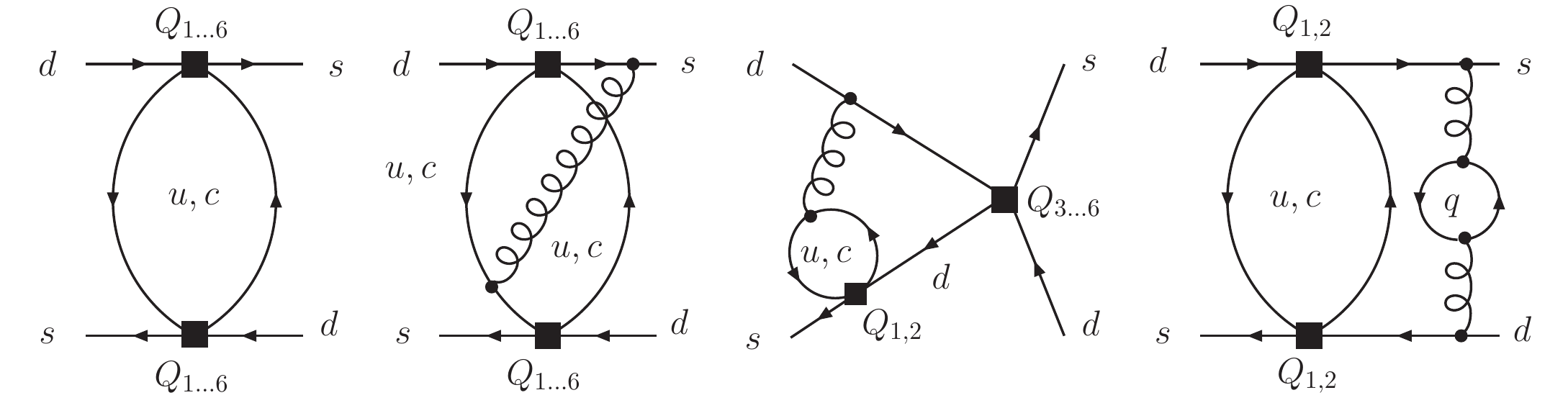}
\caption{ Sample one-, two-, and three-loop diagrams contributing to
  the NNLO mixing of dimension-six into dimension-eight operators. \label{fig:adm123l}}
\end{figure}

In order to determine the renormalisation constants, we have to
compute the divergent parts of Feynman diagrams with up to three
loops, see Figure~\ref{fig:adm123l}. We use the method suggested
in~\cite{Chetyrkin:1997fm} by Chetyrkin, Misiak and M\"unz for
extracting the UV divergences of a given Feynman diagram. The
renormalisation constants are then determined recursively by
subtracting subdivergences according to Zimmermann's forest
formula. As usual, we perform a finite renormalisation in order to
ensure the vanishing of matrix elements of evanescent operators. An
additional subtlety arises because of the presence of EOM-vanishing
operators at second order in the effective interactions: As explained
in detail in Reference~\cite{Simma:1993ky}, we have to expect
non-trivial contact terms resulting from double insertions of
$Q_{\text{eom}}$ and physical operators. We computed these terms
explicitly, showing that non-zero contributions indeed occur, and
subtracted them by an additional finite counterterm:
\begin{equation}
\hat Z_{Q_+ Q_{\text{eom}},\tilde Q_7}^{(2,0)} = \hat
Z_{Q_- Q_{\text{eom}},\tilde Q_7}^{(2,0)} = \frac{3}{8} \left( N_c - 1
- \frac{1}{N_c} - \frac{1}{N_c^2} \right) \, .
\end{equation}
This renormalisation ensures the validity of the equations of motion
also at second order in the effective interactions.

Let us now look at the equations~\eqref{eq:rgeC7} in more detail. It
turns out that these equations are equivalent to the following system
of eight equations~\cite{Herrlich:1996vf}
\begin{equation}\label{eq:eqD}
\dlogmu D = \gamma^T D \, ,
\end{equation}
where the anomalous dimension matrix and the Wilson coefficients are
now given by
\begin{equation}
  \gamma^T = 
  \begin{pmatrix}
    \gamma_Q^T&0&0\\
    \tilde{\gamma}^T_{+,7}&\tilde{\gamma}_{77}-\gamma_+&0\\
    \tilde{\gamma}^T_{-,7}&0&\tilde{\gamma}_{77}-\gamma_-
  \end{pmatrix}
  \,  , \qquad
  D(\mu) = 
  \begin{pmatrix}
    C(\mu)\\
    \tilde{C}^+_7(\mu) / C_+(\mu)\\
    \tilde{C}^-_7(\mu) / C_-(\mu)
  \end{pmatrix}
  \, ,
\end{equation}
if we decompose the Wilson coefficient $\tilde{C}_7$ as
\begin{equation}
\tilde{C}_7(\mu) = \tilde{C}_7^{+}(\mu) + \tilde{C}_7^{-}(\mu) \, .
\end{equation}
This decomposition is completely arbitrary and preserved by the
renormalisation group evolution. For instance, we may choose
$\tilde{C}_7^{+}(\mu) = \tilde{C}_7(\mu)$ and $\tilde{C}_7^{-}(\mu) =
0$. The advantage of \eqref{eq:eqD} is that it has the form of a
renormalisation group equation for a single operator insertion, and we
can use the well known explicit solution (see, for
instance, Reference~\cite{Gorbahn:2004my}). 

We obtain the anomalous dimension matrix $\gamma_Q$ of the operators
$Q_{+}, Q_{-}, Q_3 , \ldots , Q_6$ from
Reference~\cite{Gorbahn:2004my} by the basis transformation described
in the appendix and find
\begin{equation}
\gamma_Q^{(0)} = 
\begin{pmatrix}
4 & 0 & 0 & \frac{2}{3} & 0 & 0 \\[1mm]
 0 & -8 & 0 & \frac{2}{3} & 0 & 0 \\[1mm]
 0 & 0 & 0 & -\frac{52}{3} & 0 & 2 \\[1mm]
 0 & 0 & -\frac{40}{9} & \frac{4}{3}f-\frac{160}{9} & \frac{4}{9} & \frac{5}{6} \\[1mm]
 0 & 0 & 0 & -\frac{256}{3} & 0 & 20 \\[1mm]
 0 & 0 & -\frac{256}{9} & \frac{40}{3}f-\frac{544}{9} & \frac{40}{9} & -\frac{2}{3}
\end{pmatrix}\, ,
\end{equation}
\begin{equation}
\gamma_Q^{(1)} = 
\begin{pmatrix}
{\scriptstyle \frac{4}{9}{f}-7 } & {\scriptstyle 0 } & {\scriptstyle
  -\frac{748}{81} } & {\scriptstyle \frac{415}{81} } & {\scriptstyle
  \frac{82}{81} } & {\scriptstyle \frac{35}{54}} \\[1mm]
{\scriptstyle  0 } & {\scriptstyle -\frac{8}{9}{f}-14 } & {\scriptstyle
  \frac{332}{81} } & {\scriptstyle \frac{793}{81} } & {\scriptstyle
  -\frac{26}{81}} & {\scriptstyle \frac{35}{54}} \\[1mm]
{\scriptstyle  0 } & {\scriptstyle 0 } & {\scriptstyle -\frac{4468}{81} }
& {\scriptstyle -\frac{52}{9}{f}-\frac{29129}{81} } & {\scriptstyle
  \frac{400}{81}} & {\scriptstyle \frac{3493}{108}-\frac{2}{9}{f}} \\[1mm]
{\scriptstyle  0 } & {\scriptstyle 0 } & {\scriptstyle
  \frac{368}{81}{f}-\frac{13678}{243} } & {\scriptstyle \frac{1334 
   }{81}{f}-\frac{79409}{243} } & {\scriptstyle
   \frac{509}{486}-\frac{8}{81}{f} } & {\scriptstyle
   \frac{13499}{648}-\frac{5}{27}{f}} \\[1mm] 
{\scriptstyle  0 } & {\scriptstyle 0 } & {\scriptstyle
  -\frac{160}{9}{f}-\frac{244480}{81} } & {\scriptstyle -\frac{2200 
   }{9}{f}-\frac{29648}{81} } & {\scriptstyle \frac{16
     \text{Nf}}{9}+\frac{23116}{81} } & {\scriptstyle \frac{148 
   }{9}{f}+\frac{3886}{27}} \\[1mm]
{\scriptstyle  0 } & {\scriptstyle 0 } & {\scriptstyle
  \frac{77600}{243}-\frac{1264}{81}{f} } & {\scriptstyle \frac{164 
   }{81}{f}-\frac{28808}{243} } & {\scriptstyle
   \frac{400}{81}{f}-\frac{20324}{243} } & {\scriptstyle \frac{622 
   }{27}{f}-\frac{21211}{162}}
\end{pmatrix}\, ,
\end{equation}
\begin{equation}\begin{split}
\gamma_Q^{(2)} = 
\left(
\begin{matrix}
{\scriptstyle \frac{275267}{150} - \frac{260}{81} {f}^2 - \frac{52891 }{675}{f} -
  \left(\frac{160}{3}{f} + 672\right) \zeta_3 } & {\scriptstyle 0} \\[2mm]
{\scriptstyle   0 } & {\scriptstyle \frac{12297}{25} + \frac{520}{81}{f}^2-\frac{62686}{675}{f}
  +\left(\frac{320}{3}{f}+672\right) \zeta_3} \\
{\scriptstyle   0 } & {\scriptstyle 0} \\
{\scriptstyle   0 } & {\scriptstyle 0} \\
{\scriptstyle   0 } & {\scriptstyle 0} \\
{\scriptstyle   0 } & {\scriptstyle 0}
\end{matrix}
\right.
\\[4mm]
\begin{matrix}
{\scriptstyle \frac{54821}{4374} - \frac{160}{243}{f}+\frac{1360}{27} \zeta_3 } & {\scriptstyle
-\frac{8226427}{109350} -\frac{18845
   }{1458}{f}-\frac{2104}{27} \zeta_3} \\[2mm]
{\scriptstyle \frac{1064}{243}{f}+\frac{1360}{27} \zeta_3-\frac{25531}{4374} } & {\scriptstyle -\frac{26513
   }{1458}{f}-\frac{664}{27} \zeta_3+\frac{57546991}{218700}} \\[2mm]
{\scriptstyle \frac{14012}{243}{f}-\frac{608}{27} \zeta_3-\frac{4203068}{2187} } & {\scriptstyle
   \frac{272}{27}{f}^2+\frac{888605}{2916}{f}+\left(160
   {f}+\frac{39824}{27}\right) \zeta_3-\frac{18422762}{2187}} \\[2mm]
{\scriptstyle \frac{472}{81}{f}^2+\frac{217892}{2187}{f}+\left(\frac{1360
   }{9}{f}+\frac{27520}{81}\right) \zeta_3-\frac{5875184}{6561} } & {\scriptstyle -\frac{4010
   }{729}{f}^2+\frac{8860733}{17496}{f}+\left(\frac{2512
   }{27}{f}+\frac{16592}{81}\right) \zeta_3-\frac{70274587}{13122}} \\[2mm]
{\scriptstyle -\frac{2144}{81}{f}^2+\frac{358672}{81}{f}+\frac{87040}{27}\zeta_3
-\frac{194951552}{2187} } & {\scriptstyle \frac{3088}{27}{f}-\frac{2949616 
   }{729}{f}+\left(640 {f}+\frac{238016}{27}\right) \zeta_3
   -\frac{130500332}{2187}} \\[2mm]
{\scriptstyle \frac{17920}{243}{f}^2-\frac{2535466}{2187}{f}+\left(\frac{12160
   }{9}{f}+\frac{174208}{81}\right) \zeta_3+\frac{162733912}{6561} } & {\scriptstyle -\frac{159548
   }{729}{f}^2-\frac{1826023}{4374}{f}-\left(\frac{9440
   }{27}{f}+\frac{24832}{81}\right) \zeta_3+\frac{13286236}{6561}} 
\end{matrix}
\\[4mm]
\left.
\begin{matrix}
{\scriptstyle   \frac{112}{243}{f}-\frac{124}{27} \zeta_3-\frac{113417}{17496} } & {\scriptstyle
  -\frac{35
  }{324}{f}-\frac{40}{9} \zeta_3+\frac{479581}{23328}} \\[2mm]
{\scriptstyle   -\frac{140}{243}{f}-\frac{124}{27} \zeta_3+\frac{79687}{17496} } & {\scriptstyle
  -\frac{35
  }{324}{f}-\frac{70}{9} \zeta_3+\frac{242737}{23328}} \\[2mm]
{\scriptstyle   -\frac{1352}{243}{f}-\frac{496}{27} \zeta_3+\frac{674281}{4374} } & {\scriptstyle
  \frac{9284531}{11664} - \frac{26
  }{27}{f}^2-\frac{2798}{81}{f}-\left(20{f}+\frac{1921}{9}\right)
  \zeta_3} \\[2mm]
{\scriptstyle   -\frac{52}{81}{f}^2-\frac{31175}{8748}{f}-\left(\frac{136
      {f}}{9}+\frac{3154}{81}\right) \zeta_3+\frac{2951809}{52488}} & {\scriptstyle
  \frac{3227801}{8748}-\frac{65
  }{54}{f}^2-\frac{105293}{11664}{f}+\left(\frac{200}{27}-\frac{220
    }{9}{f}\right) \zeta_3} \\[2mm]
{\scriptstyle   \frac{272}{81}{f}^2-\frac{27428 }{81}{f}-\frac{13984}{27}
  \zeta_3+\frac{14732222}{2187} } & {\scriptstyle \frac{16521659}{2916} -\frac{316
  }{27}{f}^2+\frac{8081}{54}{f}-\left(200 {f}+\frac{22420}{9}\right)
  \zeta_3} \\[2mm]
{\scriptstyle   -\frac{1720}{243}{f}^2+\frac{395783}{4374}{f}+\left(-\frac{1360
    }{9}{f}-\frac{33832}{81}\right) \zeta_3-\frac{22191107}{13122} } & {\scriptstyle
  -\frac{533
  }{81}{f}^2+\frac{3353393}{5832}{f}+\left(\frac{9248}{27}-\frac{1120
    }{9}{f}\right) \zeta_3-\frac{32043361}{8748}}
\end{matrix}
\right)
\, .
\end{split}\end{equation}
Here and in the following, $f$ is the number of active quark
flavours. 

We denote the anomalous dimension for the double insertion of either
$Q_+$ or $Q_-$ and one of the operators $Q_{+}, Q_{-}, Q_3, \ldots ,
Q_6$ by
\begin{equation}
\tilde{\gamma}^T_{\pm,7} = (\tilde{\gamma}_{\pm +,7},
\tilde{\gamma}_{\pm -,7}, \tilde{\gamma}_{\pm 3,7}, \tilde{\gamma}_{\pm
  4,7}, \tilde{\gamma}_{\pm 5,7}, \tilde{\gamma}_{\pm 6,7}) \, , 
\end{equation}
and find
\begin{equation}
\tilde{\gamma}^{T(0)}_{+,7} = \left( -3, 1, 0, 0, -96, -8 \right)\, , \qquad
\tilde{\gamma}^{T(0)}_{-,7} = \left( 1, -1, 0, 0, 48, -8 \right)\, ,
\end{equation}
\begin{equation}\begin{split}
\tilde{\gamma}^{T(1)}_{+,7} &= \left( -30 , 23 , -\frac{140}{3} , -\frac{341}{9} , -\frac{248}{3} , \frac{1252}{9} \right)\, , \\ 
\tilde{\gamma}^{T(1)}_{-,7} &= \left( 23 , -46 , \frac{4}{3} , -\frac{101}{9} , -\frac{680}{3} , -\frac{164}{9} \right)\, ,
\end{split}\end{equation}
\begin{equation}\begin{split}
    \tilde{\gamma}^{T(2)}_{+,7} &= \bigg( \frac{5437543}{2808} -
    \frac{158279}{1950} {f} +252 \zeta_3 ,
    \frac{166441}{5850} {f} + \frac{106 \zeta_3}{3}-\frac{8107577}{7020}, \\
    &\quad \frac{40}{9}{f}-\frac{472 }{3}\zeta_3+\frac{27909247}{7020}
    , \frac{578}{27}{f}-\frac{2698}{9}\zeta_3+\frac{5333399}{3240} , \\
    &\quad \frac{225176}{195}{f}+\frac{6128
    }{3}\zeta_3-\frac{9973214}{1755} , \frac{4712717
    }{1755}{f}+\frac{4856}{9}\zeta_3-\frac{832816243}{10530} \bigg)\, , \\
    \tilde{\gamma}^{T(2)}_{-,7} &= \bigg( \frac{166441
    }{5850}{f}+\frac{106}{3}\zeta_3-\frac{8107577}{7020} , \frac{93707
    }{5850}{f}+\frac{104}{3}\zeta_3-\frac{23496713}{70200}, \\
    &\quad -\frac{32}{9}{f}+\frac{200
    }{3}\zeta_3-\frac{30781813}{35100} , -\frac{94
    }{27}{f}-\frac{922}{9}\zeta_3-\frac{31831601}{210600} , \\
    &\quad \frac{364552}{975}{f}+\frac{1328
    }{3}\zeta_3-\frac{83770148}{1755} , \frac{1412938999}{52650}
    -\frac{6223223}{8775}{f}+\frac{4328}{9}\zeta_3 \bigg) \, ,
\end{split}\end{equation}
at LO, NLO, and NNLO, respectively. The LO and NLO results agree with
the literature~\cite{Herrlich:1996vf} after the corresponding change
of the operator basis, described in the appendix. The NNLO
result is new. 

\begin{figure}[t]
\centering
\includegraphics[width=\textwidth]{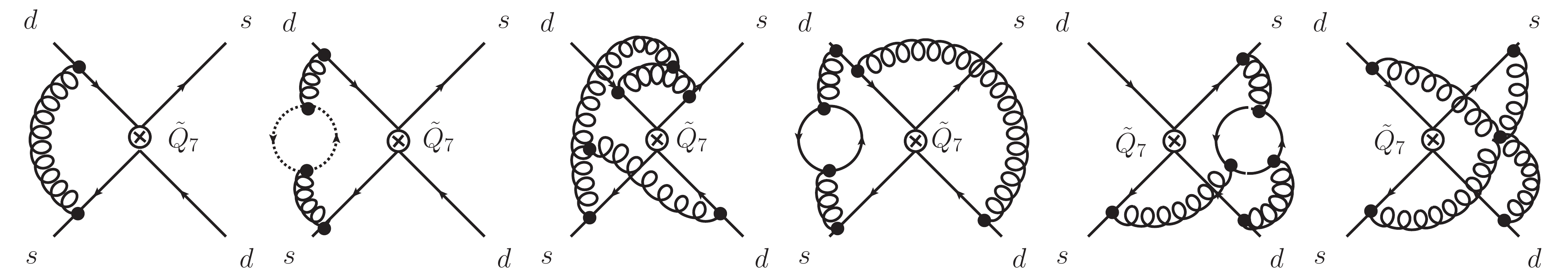}
\caption{ Sample one-, two-, and three-loop diagrams, whose divergent
  parts contribute to the anomalous dimensions of the operator $\tilde
  Q_7$. Curly lines denote gluons, dotted lines denote ghosts, and
  solid lines denote quarks. \label{fig:dim8adm}}
\end{figure}

In the calculation of $\gamma_{S2}$ (cf. the diagrams in
Figure~\ref{fig:dim8adm}) we have chosen the evanescent operators in
the dimension-eight sector in such a way that the anomalous dimension
of the operator $\tilde{Q}_{S2}$ equals the anomalous dimension of
$Q_+$ through NNLO.  Consequently $\tilde{\gamma}_{S2} = \gamma_{+}$,
and~\cite{Buras:2006gb}
\begin{equation} \label{eq:gplus} 
\begin{split}
  \gamma_{+}^{(0)} & = 4 \, , \qquad \gamma_{+}^{(1)} = \frac{4}{9} {f} - 7 \, , \\
  \gamma_{+}^{(2)} & = \frac{275\,267}{150} - \frac{52\,891}{675} {f} - \frac{260}{81}
  {f}^2 - \left(\frac{160}{3} {f} + 672\right) \zeta_3 \, .
\end{split}
\end{equation}

The explicit expressions for the QCD $\beta$ function and the
anomalous dimension of the quark mass are given by~\cite{Larin:1993tq,
  Tarasov:1982gk, Tarasov:1980au, Larin:1993tp}:
\begin{equation}\label{eq:betaqcdexp}\begin{split}
\beta_{0}&=11-\frac{2}{3}{f} \, ,\qquad
\beta_{1}=102-\frac{38}{3}{f} \, ,\qquad
\beta_{2}= \frac{2857}{2} - \frac{5033}{18} {f} + \frac{325}{54} {f}^2
\, ,
\end{split}\end{equation}
and 
\begin{equation}\label{eq:gammaexp}\begin{split}
\gamma_m^{(0)}&=8 \, ,\quad \gamma_m^{(1)} = \frac{404}{3} -
\frac{40}{9} {f} \, ,\quad \gamma_m^{(2)} = 2498 - \biggl(
\frac{4432}{27} + \frac{320}{3} \zeta_3 \biggr) {f} - \frac{280}{81} {f}^2
\, .
\end{split}\end{equation}

\subsection{Threshold Corrections at the Bottom-Quark Scale}

\begin{figure}[t]
\centering
\includegraphics[width=0.8\textwidth]{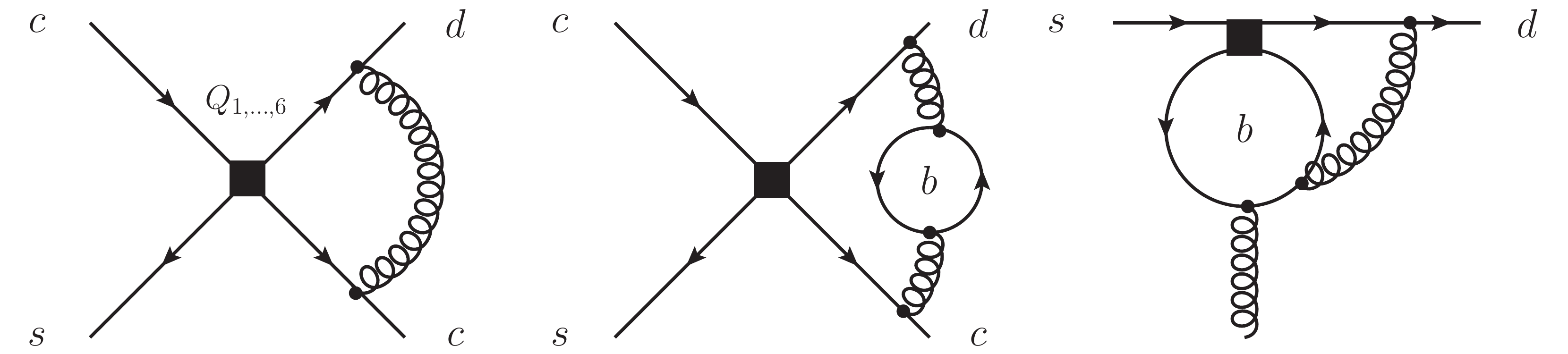}
\caption{Feynman diagrams relevant for the threshold corrections at the
  bottom quark scale. The one-loop diagram of $Q_1$ and $Q_2$ is the
  same in both theories, whereas at the two-loop level they receive
  non-trivial corrections from virtual bottom quarks. The same applies
  to insertions of the operator $\tilde Q_7$. Because the penguin
  operators mix into $Q_{\text{eom}}$, we also had to calculate
  insertions of $Q_{3,\ldots ,6}$ with one external gluon, expanding up to
  the second power in the external momenta. 
  \label{fig:nfb}}
\end{figure}

When we pass the bottom-quark threshold, we must perform a proper
matching between the effective theories with five and four
flavours. This threshold correction is computed by requiring the
equality of the Green's functions in the two theories at the matching
scale, in this case $\mu_b = \mathcal{O}(m_b)$, where $m_b$ is the
bottom-quark mass.

At NNLO, there are several sources of matching corrections. The
penguin operators are affected already at NLO, because they explicitly
depend on the number of light-quark fields. At NNLO also the matching
of the current-current and the dimension-eight operators is
non-trivial. The source of such contributions are virtual bottom
quarks in two-loop matrix elements of the form shown in
Figure~\ref{fig:nfb}. In addition, also the strong coupling constant
and the charm-quark mass are discontinuous beyond LO.

Let us write the equality of a general amplitude in the two theories
at the matching scale $\mu_f$ as
\begin{equation}\label{eq:matchthresh}
C_{f-1} (\mu_f) \langle Q_{f-1} \rangle (\mu_f) = C_{f} (\mu_f)
\langle Q_{f}  \rangle(\mu_f) \, , 
\end{equation}
the variables with subscripts $f$ and $f-1$ belonging to the $f$- and
$f-1$-flavour theory. At the bottom-quark scale, we have $f=5$. We
parameterise the matrix elements of the operators as an expansion in
the coupling constant defined in the corresponding $f$-flavour theory:
\begin{equation}
  \langle Q_{f} \rangle (\mu_f) = \left(
    1 + \frac{\alpha_s^{(f)}(\mu_f)}{4\pi} r_f^{(1)}(\mu_f) + \left(
      \frac{\alpha_s^{(f)}(\mu_f)}{4\pi} \right)^2 r_f^{(2)}(\mu_f)
  \right) \langle Q_{f} \rangle^{(0)} \, . 
\end{equation}
An additional subtlety arises, because the strong coupling constant
also gets a non-trivial matching correction at a flavour threshold. Up
to the NNLO approximation we have the relation~\cite{Wetzel:1981qg,
  Bernreuther:1981sg, Bernreuther:1983zp}
\begin{equation}\begin{split}\label{eq:asdec}
  \alpha_s^{(f)}(\mu_f) &= \alpha_s^{(f-1)}(\mu_f) \Bigg( 1 +
    \frac{\alpha_s^{(f-1)}(\mu_f)}{4\pi} \frac{2}{3}
    \log\frac{\mu_f^2}{m_f^2} \\ &- \left(
      \frac{\alpha_s^{(f-1)}(\mu_f)}{4\pi} \right)^2 \left( \frac{22}{9}
      - \frac{22}{3} \log\frac{\mu_f^2}{m_f^2} - \frac{4}{9} \log^2
      \frac{\mu_f^2}{m_f^2} \right) \Bigg) \, , 
\end{split}\end{equation}
which we use to express all quantities in terms of the coupling
constant $\alpha_s^{(f-1)}(\mu_f)$ in the effective theory with $f-1$
flavours. Here $m_f = m_f(\mu_f)$ is the \MS{} mass of the quark which
is integrated out. Note that the matching for $\tilde C_7$ starts at
order $1/\alpha_s$, so that by inverting Equation~\eqref{eq:asdec} we get a
contribution already at NLO. Similarly, we need the decoupling
relation for the charm quark mass up to NNLO \cite{Chetyrkin:1997un}:
\begin{equation}
\label{eq:decmc}
  m_c^{(f-1)}(\mu_f) = m_c^{(f)}(\mu_f) \Bigg[ 1 + \left(
      \frac{\alpha_s^{(f)}(\mu_f)}{4\pi} \right)^2 \left( \frac{89}{27}
      - \frac{20}{9} \log\frac{\mu_f^2}{m_f^2} + \frac{4}{3} \log^2
      \frac{\mu_f^2}{m_f^2} \right) \Bigg] \, . 
\end{equation}

In order to display the threshold corrections explicitly, we now
introduce the discontinuities
\begin{equation}
\delta C^{(k)}(\mu_f) = C_f^{(k)}(\mu_f) - C_{f-1}^{(k)}(\mu_f)\, , \quad
\delta r^{(k)}(\mu_f) = r_f^{(k)}(\mu_f) - r_{f-1}^{(k)}(\mu_f)\, , 
\end{equation}
of the Wilson coefficients and the matrix elements, respectively, and
find for the general solution of Equation~\eqref{eq:matchthresh}, in
case of the dimension-six Wilson coefficients:
\begin{equation}\begin{split}
\delta C^{(0)}(\mu_f) &= 0\, , \qquad \delta C^{(1)}(\mu_f) = -
C_{f}^{(0)} (\mu_f) \delta r^{(1)}(\mu_f) \, , \\
\delta C^{(2)}(\mu_f) &= - C_{f}^{(1)} (\mu_f) \left( \delta
  r^{(1)}(\mu_f) + \frac{2}{3} \log \frac{\mu_f^2}{m_f^2} \right) \\
&\quad -
C_{f}^{(0)} (\mu_f) \left( \delta r^{(2)}(\mu_f) - \delta r^{(1)}(\mu_f)
  r_{f-1}^{(1)} (\mu_f) + \frac{2}{3} r_{f}^{(1)} (\mu_f) \log
  \frac{\mu_f^2}{m_f^2} \right)\, .
\end{split}\end{equation}
Notice that the different single contributions in the last bracket may
not be finite because of spurious IR divergences, which nevertheless
cancel in the sum. The matching corrections look different for the
dimension-eight Wilson coefficients, because of the factor $1/g^2$
in front of the operator:
\begin{equation}\begin{split}
    \delta \tilde C^{(0)}(\mu_f) &= 0\, , \qquad \delta \tilde
    C^{(1)}(\mu_f) = - \tilde C_{f}^{(0)} (\mu_f) \left( \delta \tilde
      r^{(1)}(\mu_f) - \frac{2}{3}
      \log \frac{\mu_f^2}{m_f^2} \right) \, , \\
    \delta \tilde C^{(2)}(\mu_f) &= - \tilde C_{f}^{(1)} (\mu_f)
    \delta
    \tilde r^{(1)}(\mu_f) - \tilde C_{f}^{(0)} (\mu_f) \Bigg[ \delta
    \tilde r^{(2)}(\mu_f)\\ 
    &\quad - \Bigg( \delta \tilde r^{(1)}(\mu_f) - \frac{2}{3} \log
    \frac{\mu_f^2}{m_f^2} \Bigg) \tilde r_{f-1}^{(1)} (\mu_f) +
    \frac{22}{9} - \frac{22}{3} \log \frac{\mu_f^2}{m_f^2} \Bigg]\, . 
\end{split}\end{equation}
In addition, we have to take into account the terms related to the
decoupling of the charm-quark mass.

At NLO, only the matrix elements of the penguin operators get
non-vanishing contributions. They can be obtained from 
\begin{equation}
\delta r_Q^{(1)}(\mu_b) = 
\begin{pmatrix}
0 & 0 & 0 & 0 & 0 & 0 \\
 0 & 0 & 0 & 0 & 0 & 0 \\
 0 & 0 & 0 & 0 & 0 & 0 \\
 0 & 0 & 0 & - \frac{2}{3}\log\frac{\mu_b^2}{m_b^2} & 0 & 0 \\
 0 & 0 & 0 & 0 & 0 & 0 \\
 0 & 0 & 0 & 4 -\frac{20}{3}\log\frac{\mu_b^2}{m_b^2} & 0 & 0
\end{pmatrix}, 
\end{equation}
where $\delta r_Q$ denotes the difference of the matrix elements in the
subspace of dimension-six operators. At NNLO, we obtain the following
contributions for the penguin operators: 
\begin{equation}
\delta r_Q^{(2)}(\mu_b) - \delta r_Q^{(1)}(\mu_b)
  r_{Q,f=4}^{(1)} (\mu_b) + \frac{2}{3} r_{Q,f=5}^{(1)} (\mu_b) \log
  \frac{\mu_b^2}{m_b^2} = 
\begin{pmatrix}
 a_+^{(2)} & 0 & 0 & 0 & 0 & 0 \\
 0 & a_-^{(2)} & 0 & 0 & 0 & 0 \\
 0 & 0 & a_{33}^{(2)} & a_{34}^{(2)} & a_{35}^{(2)} & a_{36}^{(2)} \\
 0 & 0 & a_{43}^{(2)} & a_{44}^{(2)} & a_{45}^{(2)} & a_{46}^{(2)} \\
 0 & 0 & a_{53}^{(2)} & a_{54}^{(2)} & a_{55}^{(2)} & a_{56}^{(2)} \\
 0 & 0 & a_{63}^{(2)} & a_{64}^{(2)} & a_{65}^{(2)} & a_{66}^{(2)}
\end{pmatrix}, 
\end{equation}
where we can extract $a_+^{(2)}$ and $a_-^{(2)}$ from~\cite{Buras:2006gb} to find
\begin{equation} \label{eq:deltaCpmmub} 
    a_\pm^{(2)} = \delta r_\pm^{(2)} (\mu_b) = \mp
    \left ( 1 \mp \frac{1}{3} \right ) \left ( \frac{59}{36} + \frac{1}{3}
    L_b + L_b^2 \right ) \, 
\end{equation}
($L_b = \log \left( \mu_b^2 / m_b^2 \right)$ here and in the following
two equations). We have determined the
other entries by calculating two-loop matrix elements of the operators
$Q_+, Q_-, Q_{3\ldots 6}$ between appropriate
external states (see Figure~\ref{fig:nfb}), and find
\begin{equation}\begin{split}
    a_{33}^{(2)} & = 0 \, , \quad a_{34}^{(2)} = \frac{443}{54} -
    \frac{10}{9} L_b + \frac{10}{3} L_b^2 \, , \quad a_{35}^{(2)} = 0 \, ,\\ 
    a_{36}^{(2)} & = - \frac{85}{108} + \frac{1}{9} L_b - \frac{1}{3}
    L_b^2 \, ;\\
    a_{43}^{(2)} & = \frac{886}{243} - \frac{184}{81} L_b +
    \frac{40}{27} L_b^2 \, , \quad a_{44}^{(2)} = \frac{589}{162} -
    \frac{370}{81} L_b + \frac{37}{54} L_b^2 \, ,\\ 
    a_{45}^{(2)} & = - \frac{85}{243} + \frac{4}{81} L_b -
    \frac{4}{27} L_b^2 \, , \quad a_{46}^{(2)} = - \frac{425}{648} +
    \frac{5}{54} L_b - \frac{5}{18} L_b^2 \, ;\\ 
    a_{53}^{(2)} & = - \frac{452}{27} + \frac{80}{9} L_b \, ,\quad
    a_{54}^{(2)} = \frac{565}{27} + \frac{740}{9} L_b + \frac{100}{3}
    L_b^2 \, ,\\ 
    a_{55}^{(2)} & = \frac{38}{27} - \frac{8}{9} L_b \,
    ,\quad a_{56}^{(2)} = - \frac{383}{54} - \frac{74}{9} L_b -
    \frac{10}{3} L_b^2 \, ;\\ 
    a_{63}^{(2)} & = \frac{6874}{243} - \frac{88}{81} L_b +
    \frac{328}{27} L_b^2 \, ,\quad a_{64}^{(2)} = - \frac{2651}{162}
    + \frac{5030}{81} L_b - \frac{220}{27} L_b^2 \, ,\\
    a_{65}^{(2)} & = - \frac{826}{243} - \frac{128}{81} L_b -
    \frac{40}{27} L_b^2 \, ,\quad a_{66}^{(2)} = - \frac{467}{162} -
    \frac{266}{27} L_b - \frac{23}{18} L_b^2 \, . 
\end{split}\end{equation}
For the dimension-eight operator we find the only non-vanishing
contribution
\begin{equation}\label{eq:dr7}
\delta \tilde r_7^{(2)}(\mu_b) = 
  -\frac{59}{54} - \frac{2}{9}
  L_b - \frac{2}{3} L_b^2 = \delta r_+^{(2)} (\mu_b) \, .
\end{equation}

\subsection{Matching at the Charm-Quark Scale}

At the scale $\mu_c = \mathcal{O}(m_c)$ the charm quark is removed
from the theory as a dynamical degree of freedom, and the effective
Lagrangian is now given by Equation~\eqref{eq:lagmuc}. Requiring the
equality of the Green's functions in both theories at the charm-quark
scale leads to the matching condition
\begin{multline}
  \sum_{i,k=+,-} \sum_{j,l=+,-,3}^{6} C_i(\mu_c) C_j(\mu_c) Z_{ik}
  Z_{jl} \langle Q_{k} Q_l 
  \rangle(\mu_c) + {\tilde C}_7(\mu_c) \tilde Z_{77} \langle {\tilde Q}_7
  \rangle(\mu_c) \\ = 
  \frac{1}{32\pi^2} {\tilde C}_{S2}^{ct}(\mu_c) \tilde Z_{S2} \langle
  {\tilde Q}_{S2} \rangle(\mu_c) \, , 
\end{multline}
which we use to determine the Wilson coefficient
$\tilde{C}_{S2}^{ct}(\mu)$. To proceed, we parameterise the matrix
elements in the following way:
\begin{equation}
  \langle {\tilde Q}_7 \rangle = r_{7} \langle {\tilde Q}_{7}
  \rangle^{(0)}\, , \langle {\tilde Q}_{S2} \rangle = r_{S2} \langle
  {\tilde Q}_{S2} \rangle^{(0)} \, , \text{and} \, \langle
  Q_i Q_j \rangle (\mu_c) = \frac{m_c^2(\mu_c)}{32\pi^2}
  r_{ij,S2} \langle {\tilde Q}_{S2} \rangle^{(0)}  \, . 
\end{equation}
If we take into account the explicit factor of $m_c^2/g^2$ in the
definition of $\tilde Q_7$ and expand the Wilson coefficient
$\tilde{C}_{S2}^{ct}$ as
\begin{equation}
  \tilde{C}_{S2}^{ct}(\mu) = \frac{4\pi}{\alpha_s^{(3)}(\mu)}
  \tilde{C}_{S2}^{ct(0)}(\mu) + \tilde{C}_{S2}^{ct(1)}(\mu) +
  \frac{\alpha_s^{(3)}(\mu)}{4\pi} \tilde{C}_{S2}^{ct(2)}(\mu)\, ,  
\end{equation}
we find the following contributions to the matching: 
\begin{align}
  \tilde{C}_{S2}^{ct(0)}(\mu_c) &= 2 m_c^2(\mu_c)
  \tilde{C}_{7}^{(0)}(\mu_c) \, , \label{eq:matchmc1}\\
  \tilde{C}_{S2}^{ct(1)}(\mu_c) &= 2 m_c^2(\mu_c) \left[
    \tilde{C}_{7}^{(0)}(\mu_c) \left(r_{7}^{(1)} - r_{S2}^{(1)} -
      \frac{2}{3} \log \frac{\mu_c^2}{m_c(\mu_c)^2}\right) +
    \tilde{C}_{7}^{(1)}(\mu_c) \right]
  \notag \\
  & \quad + m_c^2(\mu_c) C_i^{(0)}(\mu_c) C_j^{(0)}(\mu_c)
  r_{ij,S2}^{(0)} \, , \label{eq:matchmc2}\\
  \tilde{C}_{S2}^{ct(2)}(\mu_c) &= 2 m_c^2(\mu_c) \bigg[
  \tilde{C}_{7}^{(0)}(\mu_c) \bigg( \delta \tilde r_7^{(2)}(\mu_c) +
  \frac{22}{9} -
  \frac{22}{3} \log \frac{\mu_c^2}{m_c(\mu_c)^2} \bigg) +
  \tilde{C}_{7}^{(2)}(\mu_c) \bigg] \notag \\ 
  & \quad + m_c^2(\mu_c) \bigg[ C_i^{(0)}(\mu_c) C_j^{(0)}(\mu_c)
  (r_{ij,S2}^{(1)} - r_{ij,S2}^{(0)}
  r_{S2}^{(1)}) \notag \\
  & \quad \hspace{1.8cm} + C_i^{(0)}(\mu_c) C_j^{(1)}(\mu_c) r_{ij,S2}^{(0)} +
  C_i^{(1)}(\mu_c) C_j^{(0)}(\mu_c) r_{ij,S2}^{(0)} \bigg] \, ,\label{eq:matchmc3}
\end{align}
where $\delta \tilde r_7^{(2)}(\mu_c)$ is given by Equation~\eqref{eq:dr7},
with $\mu_b$ and $m_b$ replaced by $\mu_c$ and $m_c$,
respectively. Notice the additional logarithms which we get by
expressing $\alpha_s^{(f=4)}$ through $\alpha_s^{(f=3)}$. These terms,
which are numerically tiny at NLO, have been neglected in
Reference~\cite{Herrlich:1996vf}.

Furthermore, we expand the charm-quark mass defined at the scale
$\mu_c$, viz. $m_c(\mu_c)$, about $m_c(m_c)$
(see~\cite{Buras:2006gb}): 
\begin{equation}\label{eq:xc}
x_c (\mu_c) = \kappa_c \left ( 1 + \frac{\alpha_s^{(4)} (\mu_c)}{4 \pi}
\xi_c^{(1)} + \left ( \frac{\alpha_s^{(4)} (\mu_c)}{4 \pi} \right )^2 \xi_c^{(2)}
\right ) x_c (m_c) \, .     
\end{equation} 
Here $\kappa_c = \eta_c^{24/25}$ with $\eta_c = \alpha_s^{(4)}
(\mu_c)/\alpha_s^{(4)} (m_c)$
and  
\begin{equation} \label{eq:xccoeff}
\begin{split}
\xi_c^{(1)} & = \frac{15212}{1875} \left ( 1 - \eta_c^{-1} \right ) \, ,
\\[2mm] 
\xi_c^{(2)} & = \frac{966966391}{10546875} - \frac{231404944}{3515625}
\eta_c^{-1} - \frac{272751559}{10546875} \eta_c^{-2}  - \frac{128}{5} \left 
( 1 - \eta_c^{-2} \right ) \zeta_3 \, .
\end{split}
\end{equation}

\begin{figure}[t]
\centering
\includegraphics[width=.6\textwidth]{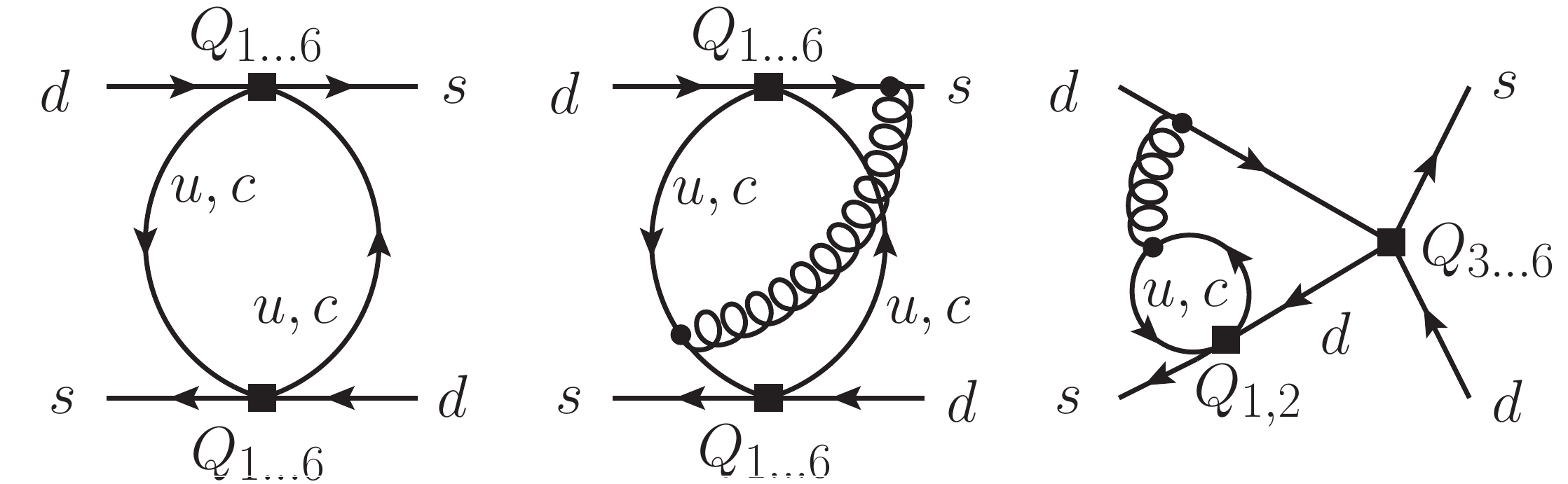}
\caption{ Sample one- and two-loop diagrams contributing to the
  matching at the charm-quark scale. \label{fig:matchmc}}
\end{figure}
In order to evaluate the equations~\eqref{eq:matchmc1},
\eqref{eq:matchmc2}, and \eqref{eq:matchmc3}, we have to
compute the finite parts of one- and two-loop Feynman diagrams of the
type shown in Figure~\ref{fig:matchmc}. In this way we find for
$r_{\pm j,S2}$ at one loop:
%
\begin{equation}
  r_{\pm j,S2}^{(0),T}(\mu_c) =  
\begin{pmatrix}
3 \log \left(\frac{\mu_c ^2}{m_c^2}\right)-\frac{3}{2} & \frac{1}{2}-\log
   \left(\frac{\mu_c ^2}{m_c^2}\right) \\
 \frac{1}{2}-\log \left(\frac{\mu_c ^2}{m_c^2}\right) & \log \left(\frac{\mu_c
   ^2}{m_c^2}\right)-\frac{1}{2} \\
 0 & 0 \\
 0 & 0 \\
 96 \log \left(\frac{\mu_c ^2}{m_c^2}\right)+224 & -48 \log \left(\frac{\mu_c
   ^2}{m_c^2}\right)-112 \\
 8 \log \left(\frac{\mu_c ^2}{m_c^2}\right)+\frac{56}{3} & 8 \log \left(\frac{\mu_c
   ^2}{m_c^2}\right)+\frac{56}{3}
\end{pmatrix}
\end{equation}
This result agrees with the one obtained in~\cite{Herrlich:1996vf}
after the appropriate basis transformation. A two-loop matching
calculation yields
%
\begin{equation}\begin{split}
  & r_{\pm j,S2}^{(1),T}(\mu_c) - r_{\pm j,S2}^{(0),T}(\mu_c)
  r_{S2}^{(1)}(\mu_c)\\ 
&=
\begin{pmatrix}
9 \log ^2\left(\frac{\mu_c ^2}{m_c^2}\right)-3 \log \left(\frac{\mu_c
   ^2}{m_c^2}\right)-\zeta_2+\frac{54926}{325} & -6 \log ^2\left(\frac{\mu_c
   ^2}{m_c^2}\right)-9 \log \left(\frac{\mu_c ^2}{m_c^2}\right)+\frac{5
   }{3}\zeta_2-\frac{89497}{1300} \\
 -6 \log ^2\left(\frac{\mu_c ^2}{m_c^2}\right)-9 \log \left(\frac{\mu_c
   ^2}{m_c^2}\right)+\frac{5}{3}\zeta_2-\frac{89497}{1300} & 9 \log
   ^2\left(\frac{\mu_c ^2}{m_c^2}\right)+29 \log \left(\frac{\mu_c
   ^2}{m_c^2}\right)+\frac{1}{3}\zeta_2+\frac{11664}{325} \\
 -4 \log ^2\left(\frac{\mu_c ^2}{m_c^2}\right)+28 \log \left(\frac{\mu_c
   ^2}{m_c^2}\right)+\frac{23}{3} & -4 \log ^2\left(\frac{\mu_c
   ^2}{m_c^2}\right)-20 \log \left(\frac{\mu_c ^2}{m_c^2}\right)-\frac{37}{3} \\
 -\frac{37}{3} \log ^2\left(\frac{\mu_c ^2}{m_c^2}\right)-\frac{59}{3} \log
   \left(\frac{\mu_c ^2}{m_c^2}\right)-\frac{1531}{36} & \frac{11}{3} \log
   ^2\left(\frac{\mu_c ^2}{m_c^2}\right)+\frac{85}{3} \log \left(\frac{\mu_c
   ^2}{m_c^2}\right)+\frac{917}{36} \\
 344 \log ^2\left(\frac{\mu_c ^2}{m_c^2}\right)+920 \log \left(\frac{\mu_c
   ^2}{m_c^2}\right)-\frac{64878}{65} & -376 \log ^2\left(\frac{\mu_c
   ^2}{m_c^2}\right)-1144 \log \left(\frac{\mu_c
   ^2}{m_c^2}\right)-\frac{964226}{325} \\
 -\frac{220}{3} \log ^2\left(\frac{\mu_c ^2}{m_c^2}\right)-\frac{1636}{3} \log
   \left(\frac{\mu_c ^2}{m_c^2}\right)-\frac{1015087}{195} & \frac{332}{3} \log
   ^2\left(\frac{\mu_c ^2}{m_c^2}\right)+\frac{1412}{3} \log \left(\frac{\mu_c
   ^2}{m_c^2}\right)+\frac{503161}{325}
\end{pmatrix}.
\end{split}\end{equation}
This result is new and completes the matching onto the three-flavour
theory. Now only a single operator contributes, and the
renormalisation group evolution below the charm-quark scale is the
same for the top-, the charm-, and the charm-top-quark contribution.

\subsection{Renormalisation Group Equations below the Charm-Quark
  Threshold}\label{sec:rgebelowcharm}

The effective Hamiltonian valid below the charm-quark threshold
contains only the single operator ${\tilde Q}_{S2}$.  The
renormalisation group evolution is now the same for the three Wilson
coefficients $\tilde{C}_{S2}^{j}$, where $j=c,t,ct$, and is described
by the evolution matrix corresponding to the anomalous dimension of
${\tilde Q}_{S2}$:
\begin{equation}\label{eq:lowev}
\tilde{C}_{S2}^{j} (\mu) = U(\mu,\mu_c) \tilde{C}_{S2}^{j} (\mu_c) \, . 
\end{equation}
By comparing~\eqref{eq:Hlo} and~\eqref{eq:lagmuc}, we see that we can
express the coefficients $\eta_{cc}$, $\eta_{tt}$, $\eta_{ct}$ as
\begin{subequations}
\label{ResEta}
\begin{eqnarray}
\!\!\!\!\!\!\!\eta_{cc} &=&
\frac{1}{m_c^2 \left(m_c\right)} \tilde C_{S2}^{(c)}\left(\mu_c\right)
\left[\alpha_s^{(3)}\left(\mu_c\right)\right]^{a_{+}}
K_{+}^{-1}(\mu_c)\, ,
\label{ResEta1}
\\
\!\!\!\!\!\!\!\eta_{tt} &=&
\frac{1}{M_W^2 S\left(x_t\left(m_t\right)\right)} 
             \tilde C_{S2}^{(t)}\left(\mu_c\right)
\left[\alpha_s^{(3)}\left(\mu_c\right)\right]^{a_{+}}
K_{+}^{-1}(\mu_c)\, ,
\label{ResEta2}
\\
\!\!\!\!\!\!\!\eta_{ct} &=& 
\frac{1}{2 M_W^2 S\left(x_c\left(m_c\right),
                      x_t\left(m_t\right)\right)} 
          \tilde C_{S2}^{(ct)}\left(\mu_c\right)
\left[\alpha_s^{(3)}\left(\mu_c\right)\right]^{a_{+}}
K_{+}^{-1}(\mu_c)\, . 
\label{ResEta3}
\end{eqnarray}
\end{subequations}
The remaining $\mu$-dependence present in \eqref{ResEta},
corresponding to the lower end of the evolution in
Equation~\eqref{eq:lowev}, is absorbed into $b\left(\mu\right)$, which
equals
\begin{eqnarray}
b\left(\mu\right) &=&
\left[\alpha_s^{(3)}\left(\mu \right)\right]^{-a_{+}} K_{+}(\mu)\, ,
\label{DefBmu}
\end{eqnarray}
where
\begin{eqnarray}
K_{+}(\mu)&=& \left(1 + J_{+}^{(1)}\frac{\alpha_s^{(3)}\left(\mu\right)}{4\pi}
  + J_{+}^{(2)} \left(\frac{\alpha_s^{(3)}\left(\mu\right)}{4\pi}\right)^2
\right)\, ,
\end{eqnarray}
and the exponent $a_{+}$ is the so-called {\em magic number} for the
operator $Q_+$ (the magic numbers as well as the matrix $J$ are
defined for instance in~\cite{Gorbahn:2004my}). This scale dependence
is cancelled by the corresponding scale dependence of the hadronic
matrix element.

\subsection{Analytical Checks of our Calculation}

Because the calculation of the NNLO contributions to $\eta_{ct}$ is
quite complex, we checked our results in several ways. 

First of all the calculation of the $\mathcal{O}(100\,000)$ Feynman
diagrams as well as the renormalisation, the computation of the
anomalous dimensions and the matching, has been performed
independently by the two of us, using a completely different setup of
computer programs. On the one hand we use qgraf \cite{Nogueira:1991ex}
for generating the diagrams; the evaluation of the integrals is then
performed using the program packages
q2e/exp/MATAD~\cite{Harlander:1997zb, Steinhauser:2000ry}, where MATAD
is written in FORM~\cite{Vermaseren:2000nd} and based on the
Integration-By-Parts
algorithm~\cite{Tkachov:1981wb,Chetyrkin:1981qh}. In addition, we have
written our own FORM routine in order to evaluate two-loop diagrams
with an arbitrary number of (possibly vanishing) masses, using the
algorithm described in~\cite{Bobeth:1999mk, Davydychev:1992mt}. On the
other hand, all calculations have been performed using an completely
independent setup, based on Feynarts \cite{Hahn:2000kx} and
Mathematica.

As a check of our calculation, we verified that all anomalous
dimensions, Wilson coefficients, and matrix elements are independent
of the gauge-fixing parameter $\xi$. Because of the complexity of the
analytical expressions, for the three-loop penguin insertions we kept
only the first power in $\xi$ for our check.

Another very useful check is the locality of the counterterms, which
is an implication of renormalisability. In a mass independent
renormalisation scheme this means that the renormalisation factors $Z$
depend on $\mu$ only through the coupling constants. We have checked
this explicitly and found $\mu$-independence of all our
renormalisation constants.

We have also checked analytically that $\eta_{ct}$ is independent of
the matching scales $\mu_W$, $\mu_b$, and $\mu_c$ to the considered
order of the strong coupling constant, by expanding the full solution
of the renormalisation group equations about the respective matching
scale.

As a cross-check, we confirm the NLO results of Herrlich and
Nierste~\cite{Herrlich:1996vf} for the first time.

\section{Discussion and Numerics}\label{sec:dnum}

In this section we present the numerical value of $\eta_{ct}$ at NNLO
and discuss the theoretical uncertainty, as well as the impact on
$\epsilon_K$. Our input parameters are collected in
Table~\ref{tab:num}.

\begin{table}[t]
\begin{center}
\begin{tabular}{|c|l|c||c|l|c|}\hline
  Parameter&Value&Ref.&
  Parameter&Value&Ref.
  \\\hline
  $M_W$&$80.399(23)$\,GeV&\cite{PDG2010}&
  $\alpha_s(M_Z)$&$0.1184(7)$&\cite{PDG2010}
  \\\hline
  $m_t(m_t)$&$163.7(1.1)$\,GeV&\cite{:1900yx}
  &$F_K$&156.1(8)\,MeV&\cite{Antonelli:2008jg}
  \\\hline
  $m_b(m_b)$&$4.163(16)$\,GeV&\cite{Chetyrkin:2009fv}&
  $G_F$&$1.166\,367(5)\times 10^{-5}\text{GeV}^{-2}$&\cite{PDG2010}
  \\\hline
  $m_c(m_c)$&$1.286(13)$\,GeV&\cite{Chetyrkin:2009fv}&
  $\lambda$&$0.2255(7)$&\cite{Antonelli:2008jg}
  \\\hline
  $M_{K}$&497.614(24)\,MeV&\cite{PDG2010}&
  $\left| V_{cb} \right|$ & $4.06(13) \times 10^{-2}$ &
  \cite{PDG2010} 
  \\\hline
  $\kappa_{\epsilon}$&0.94(2)&\cite{Buras:2010pz}&
  $M_{B_d}$&$5.2795(3)$\,GeV&\cite{PDG2010}
  \\\hline
  $\Delta M_K$&5.292(9)/ns&\cite{PDG2010}&
  $M_{B_s}$&$5.3663(6)$\,GeV&\cite{PDG2010}
  \\\hline
  $\Delta M_d$&0.507(5)/ps&\cite{PDG2010}&
  $\Delta M_s$&17.77(12)/ps&\cite{PDG2010}
  \\\hline
  $\xi_s$&1.243(28)&\cite{Laiho:2009eu}&
  $\eta_{tt}$&0.5765(65)&\cite{Buras:1990fn}
  \\\hline
  $\hat B_K$&0.725(26)&\cite{Laiho:2009eu}&
  $\eta_{cc}$&1.43(23)&\cite{Herrlich:1996vf}
  \\\hline
  $\sin 2 \beta$&0.671(23)&\cite{PDG2010}&
  &&
  \\\hline
\end{tabular}
\caption{Input parameters used in our numerical
  analysis. \label{tab:num}}
\end{center}
\end{table}

The theoretical uncertainty of $\eta_{ct}$ is related to the
truncation of the perturbation series. We estimate it by considering
the remaining scale dependence, the different methods to evaluate the
running strong coupling constant, and the size of the NNLO
corrections. Varying $\mu_c$ from 1 to 2 GeV and $\mu_W$ from 40 to
200 GeV, we find the following numerical value at NNLO,
\begin{equation}
  \eta_{ct} = 0.496 \pm 0.045 _{\mu_c} \pm 0.013 _{\mu_W} \pm 0.002
  _{\alpha_s} \pm 0.001 _{m_c} \pm 0.0002 _{m_t} \, , 
\end{equation}
where we also display the parametric uncertainties stemming from the
experimental error on $\alpha_s$, $m_c$, and $m_t$. The dependence on
the scale $\mu_b$ is completely negligible. 

\begin{figure}
\begin{minipage}[t]{0.47\textwidth}
\begin{center}
\includegraphics[width=0.9\textwidth]{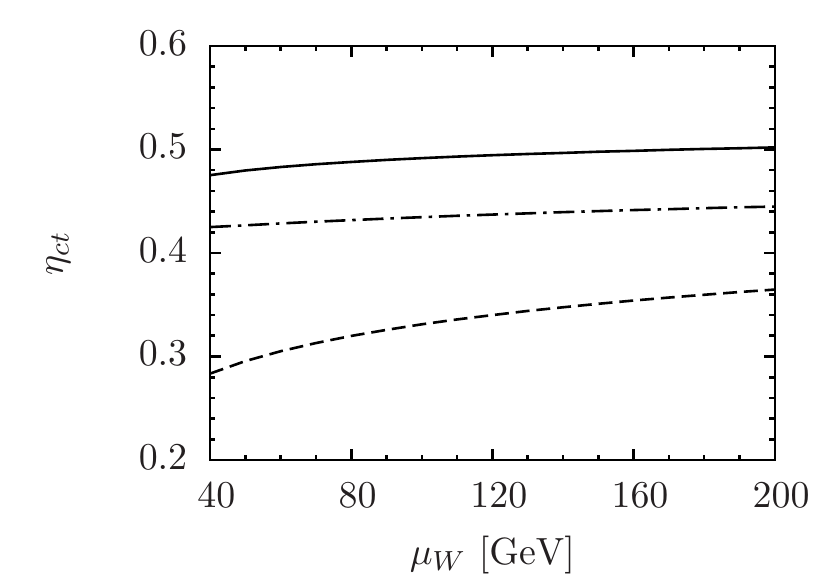}
\end{center}
\vspace{-5mm}
\caption{$\eta_{ct}$ as a function of $\mu_W$ at LO (dashed line), NLO
  (dashed-dotted line), and NNLO QCD (solid line). }
\label{fig:muwasex}
\end{minipage}
\hfill
\begin{minipage}[t]{0.47\textwidth}
\begin{center}
\includegraphics[width=0.9\textwidth]{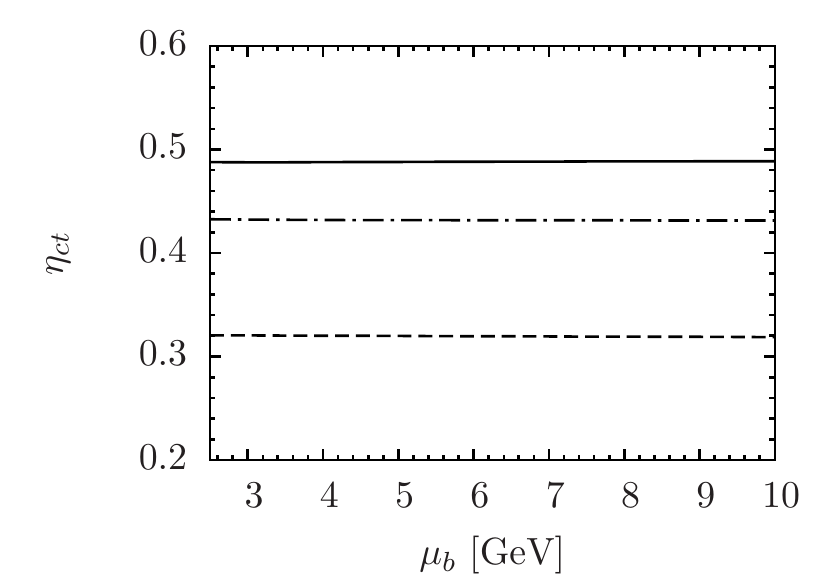}
\end{center}
\vspace{-5mm}
\caption{$\eta_{ct}$ as a function of $\mu_b$ at LO (dashed line), NLO
  (dashed-dotted line), and NNLO QCD (solid line). }
\label{fig:mubasex}
\end{minipage}
\end{figure}

\begin{figure}[h]
\begin{center}
\includegraphics[width=0.7\textwidth]{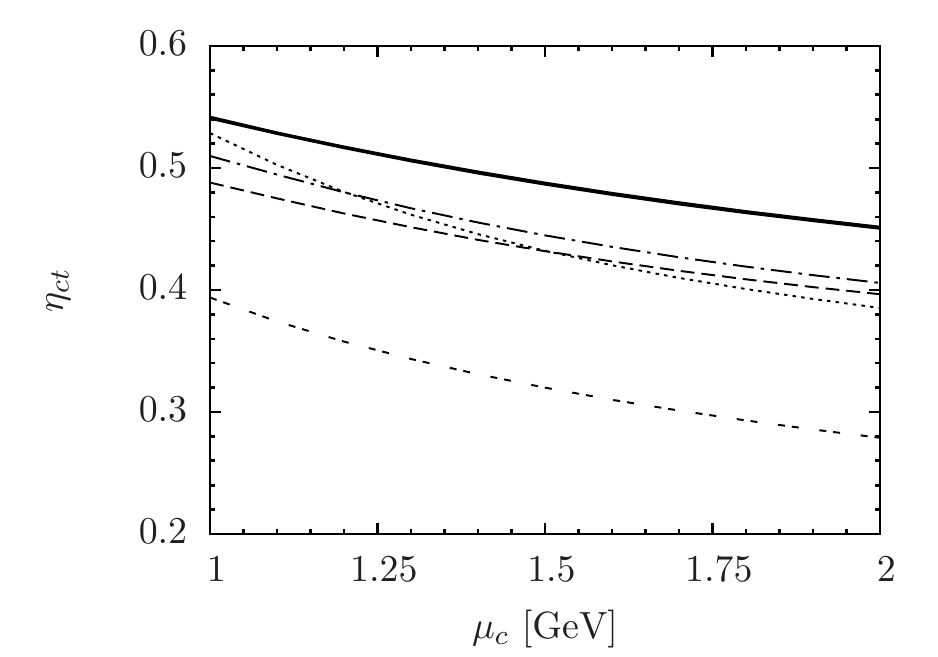}
\end{center}
\vspace{-5mm}
\caption{$\eta_{ct}$ as a function of $\mu_c$. The LO result is
  represented by the double-dotted line. The dashed, dotted, and
  dashed-dotted lines correspond the NLO value of $\eta_{ct}$, with
  $\alpha_s$ evaluated by method 1, 2, and 3, explained in the
  text. The solid lines show the corresponding NNLO results; the
  ambiguity is almost cancelled. }
\label{fig:mucasall}
\end{figure}

The dependence on the electroweak matching scale $\mu_W$ is shown in
Figure~\ref{fig:muwasex}. We have plotted $\eta_{ct}$ as a function of
$\mu_W$ in the range from 40 GeV to 200 GeV, where we fixed the other
scales as $\mu_b=5 \, \text{GeV}$ and $\mu_c=1.5 \, \text{GeV}$,
respectively. The relatively weak residual dependence on $\mu_W$ at
NLO is slightly increased at NNLO. By contrast, the dependence on
$\mu_b$, which is shown in Figure~\ref{fig:mubasex}, fixing $\mu_W=80
\, \text{GeV}$ and $\mu_c=1.5 \, \text{GeV}$, is completely
negligible.
The dependence on the scale $\mu_c$ is shown in
Figure~\ref{fig:mucasall}, where we vary $\mu_c$ in the range from 1
to 2 GeV, fixing $\mu_W=80 \, \text{GeV}$ and $\mu_b=5 \,
\text{GeV}$. In addition, we have plotted $\eta_{ct}$ corresponding to
three different possibilities of calculating $\alpha_s(\mu_c)$ from
the experimental input value of $\alpha_s(M_Z)$: One method (1) is to
solve the renormalisation group equation for $\alpha_s$
numerically. Furthermore, it is possible to compute $\alpha_s$ by
first determining the scale parameter $\Lambda_{\text{QCD}}$. This can
be achieved by using the explicit solution for $\Lambda_{\text{QCD}}$
without expansion in $\alpha_s$ (method 2) or by iteratively solving
this equation for $\Lambda_{\text{QCD}}$ and from this value
determining $\alpha_s$ (method 3). The dashed, dotted, and
dashed-dotted lines in Figure~\ref{fig:mucasall}, each of them
representing the NLO result for $\eta_{ct}$, correspond to these three
possibilities of determining $\alpha_s$, respectively. We used the
mathematica package RunDec \cite{Chetyrkin:2000yt} for the numerical
evaluation. Note that the difference between these three methods
vanishes almost entirely at NNLO.
On the other hand, at NLO the effect is sizeable and thus contributes
to the theoretical uncertainty. Varying $\mu_c$ and $\mu_W$ in the
same range as above, we find at NLO
\begin{equation}
  \eta_{ct}^{\text{NLO}} = 0.457 \pm 0.072 _{\mu_c} \pm 0.01 _{\mu_W} \pm 0.0001
  _{\alpha_s} \pm 0.002 _{m_c} \pm 0.0003 _{m_t} \, ,
\end{equation}
where there error indicated by the subscript ``$\mu_c$'' includes the
effect of the three ways of determining $\alpha_s$. For the variation
of the scale $\mu_W$ we have used only method 1 for evaluating
$\alpha_s$ in order to avoid double-counting of the related
uncertainty\footnote{Otherwise the error would amount to $\pm
  0.018_{\mu_W}$.}. Again we have included the parametric
uncertainties related to $\alpha_s$, $m_c$, and $m_t$.

The authors of Reference~\cite{Herrlich:1996vf} have varied $\mu_c$ in
the smaller range from 1.1 to 1.6 GeV, using a procedure equivalent to
method 3 above for determining $\alpha_s$. By looking at the explicit
values of $\eta_{ct}$ in Figure~\ref{fig:mucasall} we see that the two
error bands at NLO and NNLO, resulting from this smaller range of
$\mu_c$, have almost no overlap. Now, with the NNLO results at hand,
we see that our range for $\mu_c$ leads to a better estimate of the
theoretical uncertainty.

Looking at Figure~\ref{fig:mucasall}, it is striking that the scale
dependence of the NLO result is barely reduced at NNLO. In order to
understand this behaviour, let us look at the remaining $\mu_c$
dependence, which is the most pronounced, in more detail. It
originates from terms proportional to higher powers of $\alpha_s$
times logarithms of the renormalisation scale that are contained in
the explicit solutions of the renormalisation group equations. These
terms are only partially cancelled due to our truncating the
perturbative expansion of the matrix elements at the charm-quark
scale.

We have separated the contributions to $\eta_{ct}$ of the different
Wilson coefficients multiplying the matrix elements at the charm-quark
scale (cf. Equations~\eqref{eq:matchmc1}-\eqref{eq:matchmc3}). To this
end we have chosen the operator basis as in
Reference~\cite{Herrlich:1996vf}, where we use the diagonal operator
basis only in one dimension-six subspace, and $Q_1, \ldots, Q_6$ in
the other. It turns out that only one contribution, proportional to
the combination $C_- C_2$, shows a strong scale dependence. Although
the size of the individual contributions certainly depends on the
chosen renormalisation scheme, the general pattern is independent of
this convention. It is related to the vanishing of the entry in the LO
anomalous dimension tensor corresponding to the two operators $Q_-$
and $Q_2$. This incidence leads to a behaviour of the scale dependence
for this single combination which would be expected from a NLO
calculation, and dominates the scale dependence of the NNLO result.

In general, the perturbation series for the $\Delta S=2$ four-quark
operator in an effective three-flavour theory is not expected to
converge as well as, for instance, the perturbation series in
Ref.~\cite{Buras:2006gb}, where the mixing into a semileptonic
operator was calculated.

Finally we remark that the absolute value of the NNLO correction is of
the same order of magnitude as the range of $\eta_{ct}$ at NNLO in the
interval $\mu_c = 1\ldots 2 \,\, \text{GeV}$, so that using the size
of the NNLO corrections as an estimate of the theoretical uncertainty
yields approximately the same error as using the scale variation in
the quoted interval.

As a summary of the discussion above, we give the following final
estimate for the charm-top-quark contribution to $\epsilon_K$ at NNLO:
\begin{equation}
\eta_{ct} = 0.496 \pm 0.047 \, .
\end{equation}
(For comparison, an error estimate using a range for $\mu_c$ as in
Reference~\cite{Herrlich:1996vf} would yield $\eta_{ct} = 0.504 \pm
0.025$.) The parametric uncertainty is essentially negligible with
respect to the theoretical uncertainty. Compared to our NLO value,
\begin{equation}\label{eq:etactNLO}
\eta_{ct}^{\text{NLO}} = 0.457 \pm 0.073 \, ,
\end{equation}
this corresponds to a positive shift of approximately $7\%$.

Before we conclude this section, we study the impact of our
calculation on the prediction of $|\epsilon_K|$. To this end we use
the following formula\footnote{A term proportional to
  $\text{Re}\lambda_t/\text{Re}\lambda_c = \mathcal{O}(\lambda^4)$ has
  been neglected in Equation~\eqref{eq:eKformula} (see
  Reference~\cite{Buchalla:1995vs}). }~\cite{Buchalla:1995vs,Buras:2008nn}:
\begin{equation}\label{eq:eKformula}
|\epsilon_K| = \kappa_\epsilon C_\epsilon \hat B_K |V_{cb}|^2
\lambda^2 \bar \eta (|V_{cb}|^2(1-\bar\rho)\eta_{tt}S(x_t) +
\eta_{ct} S(x_c,x_t) - \eta_{cc} S(x_c)) \, ,
\end{equation}
where
\begin{equation}\label{eq:Ce}
C_\epsilon = \frac{G_F^2 F_K^2 M_{K^0} M_W^2}{6\sqrt{2}\pi^2\Delta
  M_K}\, .
\end{equation}
We write $\bar\eta = R_t \sin\beta$ and $1-\bar\rho = R_t \cos\beta$,
where $R_t$ is given by
\begin{equation}\label{eq:Rt}
R_t \approx \frac{\xi_s}{\lambda} \sqrt{\frac{M_{B_s}}{M_{B_d}}}
\sqrt{\frac{\Delta M_d}{\Delta M_s}}
\end{equation}
and $\xi_s = (F_{B_s}\sqrt{\hat B_s})/(F_{B_d}\sqrt{\hat B_d})$ is a
ratio of $B$ meson decay constants and bag factors that can be
computed on the lattice with high precision~\cite{Laiho:2009eu}. Using
the numerical values given in Table~\ref{tab:num}, we obtain
\begin{equation}
  |\epsilon_K| = (1.90\pm 0.04_{\eta_{cc}} \pm 0.02_{\eta_{tt}}
  \pm 0.07_{\eta_{ct}} \pm 0.11_{\text{LD}} \pm 0.22_{\text{parametric}})
  \times 10^{-3} \, . 
\end{equation}
The first three errors correspond to $\eta_{cc}$, $\eta_{tt}$,
$\eta_{ct}$, respectively. The error indicated by LD originates from
the long-distance contribution, namely $\xi_s$, $\hat B_K$, and
$\kappa_\epsilon$, which account for $40\%$, $37\%$, and $22\%$ of the
long-distance error, respectively. The main share of the parametric
error stems from $|V_{cb}|$ $(59\%)$ and $\sin(2\beta)$ $(19\%)$, while
all other contributions are well below $10\%$. All errors have been
added in quadrature.

Compared to the prediction using the NLO value $\eta_{ct}^{\text{NLO}}$,
\begin{equation}
  |\epsilon_K^{\text{NLO}}| = (1.83 \pm 0.04_{\eta_{cc}} \pm 0.02_{\eta_{tt}}
  \pm 0.11_{\eta_{ct}}  \pm 0.10_{\text{LD}} \pm
  0.22_{\text{parametric}}) \times 10^{-3} \, ,  
\end{equation}
this corresponds to a shift by approximately $3 \%$.

\section{Conclusion}

We have performed a complete NNLO QCD analysis of the charm-top-quark
contribution $\eta_{ct}$ to the $|\Delta S|=2$ effective Hamiltonian
$\mathcal{H}_{f=3}^{|\Delta S|=2}$. We confirm the analytical results
for $\eta_{ct}$ obtained at NLO in Reference~\cite{Herrlich:1996vf}
for the first time. 

Some of our results are useful beyond $\eta_{ct}$. The anomalous
dimension of the operator $\tilde Q_{S2}$ can be employed to compute
the large NNLO logarithms of $B^0$ -- $\bar B^0$ mixing and comprise
part of a NNLO calculation of $\eta_{tt}$. The NNLO matching
corrections at the bottom-quark threshold have further applications in
Kaon physics.

Our numerical results for $\eta_{ct}$ can be summarised by a $7 \%$
positive shift in the NNLO prediction with respect to the NLO value,
leading to $\eta_{ct} = 0.496 \pm 0.047$. This corresponds to an
enhancement of $\epsilon_K$ by roughly $3 \%$, yielding
$|\epsilon_K| = (1.90 \pm 0.26) \times 10^{-3}$. With our calculation
we solidified the theory prediction of $\epsilon_K$, strengthening its
role as an important constraint for models of new physics.

\subsection*{Acknowledgements}

JB would like to thank Ulrich Nierste for suggesting the topic,
Matthias Steinhauser for providing us with an updated version of
MATAD, and Emmanuel Stamou for his help in improving our FORM
routines. We also thank Ulrich Nierste for constant encouragement and
discussions, and Andrzej Buras, Ulrich Nierste, Guido Bell and
Emmanuel Stamou for their careful reading of the manuscript.

This research was supported by the DFG cluster of excellence ``Origin
and Structure of the Universe''. The work of JB was supported by the
EU Marie-Curie grant MIRG--CT--2005--029152, BMBF grant 05 HT6VKB, and
by the DFG--funded ``Graduiertenkolleg Hochenergiephysik und
Teilchenastrophysik'' at the University of Karls\-ruhe. MG thanks the
Galileo Galilei Institute for Theoretical Physics for the hospitality
and the INFN for partial support during the completion of this work.

\appendix

\section{Change of the Operator Basis}\label{ch:chb}

In this appendix we examine how the Wilson coefficients and the
anomalous dimensions transform under a change of the operator basis.
This is important for two reasons: In order to find a compact form for
the renormalisation group equations for double operator insertions, we
have seen it to be useful to work in a diagonal operator basis in the
subspace of current-current operators. However, the calculation of the
dimension-six anomalous dimensions and Wilson coefficients has been
performed in the literature in the basis given
in~\cite{Chetyrkin:1997gb, Gorbahn:2004my}. Moreover, we had to
transform our results in order to compare them with results that are
available in the literature and have been calculated using yet another
operator basis~\cite{Herrlich:1996vf}.

As is well known, a general change of the operator basis consists of a
linear transformation and a corresponding change of the
renormalisation scheme~\cite{Gorbahn:2004my}. Therefore, let us first
as a preparation derive the transformation properties of the anomalous
dimensions for an arbitrary change of the renormalisation scheme. This
generalises the already known results. Suppose we perform the
following change of scheme for the Wilson coefficients
\begin{align}
  C_i \to C'_i &= C_j \rho_{ji}^{-1} \, , \label{eq:tw6}\\
  \tilde{C}_k \to \tilde{C}'_k &= \tilde{C}_j \tilde{\rho}_{jk}^{-1} -
  C_l C_m \hat{\rho}_{lm,k} \, .\label{eq:tw8}
\end{align}
As before, we have denoted Wilson coefficients belonging to
dimension-eight operators with a tilde and those belonging to
dimension-six operators without superscript. Furthermore, we
introduced the parameters $\rho$, $\tilde\rho$ and $\hat\rho$, which
parameterise the finite transformations:
\begin{align}
  \rho_{ij} &= \delta_{ij} - \frac{\alpha_s}{4\pi}\rho_{ij}^{(1)} -
  \left(\frac{\alpha_s}{4\pi}\right)^2 \left( \rho_{ij}^{(2)} -
    \rho_{ik}^{(1)}\rho_{kj}^{(1)} \right) + \mathcal{O}(\alpha_s^3) \, , \\
  \tilde\rho_{ij} &= \delta_{ij} -
  \frac{\alpha_s}{4\pi}\tilde{\rho}_{ij}^{(1)} -
  \left(\frac{\alpha_s}{4\pi}\right)^2 \left( \tilde{\rho}_{ij}^{(2)} -
    \tilde{\rho}_{ik}^{(1)} \tilde{\rho}_{kj}^{(1)} \right) +
  \mathcal{O}(\alpha_s^3) \, , \\ 
  \hat{\rho}_{lm,k} &= \frac{\alpha_s}{4\pi} \hat{\rho}_{lm,k}^{(1)} +
  \left(\frac{\alpha_s}{4\pi}\right)^2 \hat{\rho}_{lm,k}^{(2)} +
  \mathcal{O}(\alpha_s^3) \, .
\end{align}
Then, in order for the effective Hamiltonian of the form
\begin{equation}
H_{\text{eff}} = C_i Z_{ij} Q_j + \bigl( \tilde C_i \tilde Z_{ik} +
C_i C_j \hat Z_{ij,k} \bigr) \tilde Q_k 
\end{equation}
to stay invariant, the renormalisation constants must transform as
\begin{align}
  Z_{ij} \to Z'_{ij} &= \rho_{ik} Z_{kj} \, , \\
  \tilde{Z}_{ij} \to \tilde{Z}'_{ij} &= \tilde{\rho}_{ik} \tilde{Z}_{kj}
  \, , \\ 
  \hat{Z}_{ij,k} \to \hat{Z}'_{ij,k} &= \rho_{il}\rho_{jm}\hat{Z}_{lm,k}
  + \rho_{il}\rho_{jm}\hat{\rho}_{lm,p}\tilde{\rho}_{pq}\tilde{Z}_{qk}\, . 
\end{align}

The transformation of the anomalous dimensions can now be obtained by
inserting the transformed renormalisation constants into the defining
equation for the anomalous dimension matrix~\eqref{eq:gammatdef}, and
the anomalous dimension tensor~\eqref{eq:adtdef}, respectively. In
this way we obtain the well-known results for the case of single
insertions~\cite{Gorbahn:2004my, Buras:1991jm, Ciuchini:1993vr}:
\begin{align}
  \gamma'^{(0)} &= \gamma^{(0)} \, , \label{eq:chgm1}\\
  \gamma'^{(1)} &= \gamma^{(1)} - [\rho^{(1)}, \gamma^{(0)}] -2\beta_0
  \rho^{(1)} \, , \label{eq:chgm2}\\
  \gamma'^{(2)} &= \gamma^{(2)} - [\rho^{(2)}, \gamma^{(0)}] -
  [\rho^{(1)}, \gamma^{(1)}] + \rho^{(1)} [\rho^{(1)}, \gamma^{(0)}]
  \notag \\ 
  &\quad - 4\beta_0\rho^{(2)} - 2\beta_1\rho^{(1)} +
  2\beta_0\rho^{(1)}\rho^{(1)} \, . \label{eq:chgm3} 
\end{align}
The general transformation law for the anomalous dimension tensor for
double insertion reads:\footnote{Note that additional finite
  contributions arise if we include the factor of $m_c^2/g^2$ in the
  definition of the dimension-eight operators. }
\begin{align}
  \hat\gamma'^{(0)}_{ij,k} &= \gamma_{ij,k}^{(0)} \, , \label{eq:chgtf1}\\
  \hat\gamma'^{(1)}_{ij,k} &= \gamma^{(1)}_{ij,k} +
  \hat{\rho}_{ij,l}^{(1)}\tilde{\gamma}_{lk}^{(0)} +
  2\hat{\rho}_{ij,k}^{(1)}\beta_0 +
  \hat{\gamma}_{ij,l}^{(0)}\tilde{\rho}_{lk}^{(1)} \notag \\
  &\quad - \left\{ \gamma_{il}^{(0)}\delta_{jm} +
    \delta_{il}\gamma_{jm}^{(0)} \right\}\hat{\rho}_{lm,k}^{(1)} -
  \left\{ \rho_{il}^{(1)}\delta_{jm} +
    \delta_{il}\rho_{jm}^{(1)} \right\}\hat{\gamma}_{lm,k}^{(0)} \, . \label{eq:chgtf2}
\end{align}

Let us now examine how the anomalous dimensions and the Wilson
coefficients change under a basis transformation. In four space-time
dimensions, a change of $n$ dimension-six operators $Q$ and $m$
dimension-eight operators $\tilde Q$ is simply given by a linear
transformation
\begin{equation}\label{eq:btrafo}
  Q_i \to Q'_i = R_{ij}Q_j\, , \qquad \tilde Q_i \to \tilde Q'_i =
  \tilde R_{ij}\tilde Q_j \, , 
\end{equation}
described by matrices $R\in \text{GL}(n)$, $\tilde R\in \text{GL}(m)$.
Under this transformation the renormalisation constants change
according to
\begin{equation}\label{eq:ztrafo}
  Z'_{ij} = R_{ik}Z_{kl}R_{lj}^{-1} \, , \qquad \tilde Z'_{ij} = \tilde
  R_{ik}\tilde Z_{kl}\tilde R_{lj}^{-1} \, , 
  \qquad \hat{Z}'_{kn,l} =
  R_{kk'}R_{nn'}\hat{Z}_{k'n',l'}\tilde{R}_{l'l}^{-1} \, .
\end{equation}

In general, the situation is more complicated because of the presence
of evanescent operators. As explained in detail in
Reference~\cite{Gorbahn:2004my}, a change of the operator basis
consists of a linear transformation and a finite renormalisation; the
latter is needed in order to restore the standard
$\overline{\text{MS}}$ definition of the renormalisation constants. 

We can write a general transformation among all dimension-six operators
as
\begin{equation}\label{eq:gentmat}
\begin{pmatrix}
Q'\\
E'
\end{pmatrix}
=
\begin{pmatrix}
R&0\\
0&M
\end{pmatrix}
\begin{pmatrix}
1&0\\
\epsilon U + \epsilon^2 V&1
\end{pmatrix}
\begin{pmatrix}
1&W\\
0&1
\end{pmatrix}
\begin{pmatrix}
Q\\
E
\end{pmatrix}\, ,
\end{equation}
where the matrices $R$ and $M$ parameterise a linear transformation
among the physical and evanescent operators $Q$ and $E$, respectively,
$W$ parameterises the addition of multiples of evanescent operators to
the physical operators, and $U$ and $V$ parameterise the addition of
multiples of $\epsilon$ and $\epsilon^2$ times physical operators to
the evanescent operators, respectively. We apply a transformation of
the same form to the dimension-eight operators, where we denote the
corresponding matrices by a tilde, as before. The finite
renormalisation constants can now be determined by requiring that an
effective amplitude of the form $C_iZ_{ij}\langle Q_j \rangle +
(\tilde C_l \tilde Z_{lk} + C_iC_j \hat Z_{ij,k})\langle \tilde Q_k
\rangle$ be invariant under the basis transformation and be
renormalised according to the \MS{} prescription.

Let us start with the anomalous dimension matrices for the mixing of
dimension-six into dimension-six operators. The finite renormalisation
induced by the change~\eqref{eq:gentmat} is given
by~\cite{Gorbahn:2004my,Buras:2006gb}
\begin{align}\label{eq:finz}
  Z_{QQ}^{'(1,0)} &= R \left[ W Z_{EQ}^{(1,0)} - \left( Z_{QE}^{(1,1)} +
      W Z_{EE}^{(1,1)} - \frac{1}{2} \gamma^{(0)} W \right) U \right]
  R^{-1}\, , \notag\\ 
  Z_{QQ}^{'(2,0)} &= - R \left ( Z^{(2,1)}_{QE} U + Z^{(2,2)}_{QE} V -
    \frac{1}{2} Z^{(1,1)}_{QE} V \gamma^{(0)} \right ) R^{-1} \, ,
\end{align}
where
\begin{equation}
  Z^{(2,2)}_{QE} = \frac{1}{2} \left ( Z^{(1,1)}_{QE} Z^{(1,1)}_{EE} +
    \frac{1}{2} \gamma^{(0)} Z^{(1,1)}_{QE} - \beta_0 Z^{(1,1)}_{QE}
  \right ) \, .
\end{equation}
We have set $W$ to zero in the second line of Equation~\eqref{eq:finz}
as these terms are not needed in our work. We now find the
transformation law for the anomalous dimension matrices in a
straightforward manner using Equations~\eqref{eq:chgm1}
to~\eqref{eq:chgm3}:
\begin{align}\label{eq:admtrafo}
\gamma^{'(0)} &= R\gamma^{(0)}R^{-1}\, , \notag\\
\gamma^{'(1)} &= R\gamma^{(1)}R^{-1} - \left[
  Z_{QQ}^{'(1,0)},\gamma^{'(0)} \right] - 2 \beta_0 Z_{QQ}^{'(1,0)}\, , 
\notag\\
\gamma^{'(2)} &= R\gamma^{(2)}R^{-1} - \left[
  Z_{QQ}^{'(2,0)},\gamma^{'(0)} \right] - \left[
  Z_{QQ}^{'(1,0)},\gamma^{'(1)} \right] + \left[
  Z_{QQ}^{'(1,0)},\gamma^{'(0)} \right] Z_{QQ}^{'(1,0)} \notag\\
 &\quad - 4 \beta_0 Z_{QQ}^{'(2,0)} - 2 \beta_1 Z_{QQ}^{'(1,0)} + 2 \beta_0
 \left( Z_{QQ}^{'(1,0)} \right)^2\, .
\end{align}
The Wilson coefficients change according to
\begin{equation}\label{eq:cprime}
  C'(\mu) = \left [ 1 + \frac{\alpha_s (\mu)}{4 \pi} Z'^{(1,0)}_{QQ} +
    \left ( \frac{\alpha_s (\mu)}{4 \pi} \right )^2 Z'^{(2,0)}_{QQ}
  \right ]^T \! \! ( R^{-1} \big )^T C (\mu) \, .
\end{equation} 

Clearly, the transformation law of the anomalous dimension matrix
describing the mixing among the dimension-eight operators themselves
is given by a formula completely analoguous to~\eqref{eq:admtrafo}. In
order to find the transformation law of the anomalous dimension
tensor, describing the mixing of dimension-six into dimension-eight
operators, and of the dimension-eight Wilson coefficients, we apply
the same method as above. In addition to the finite renormalisation
constants~\eqref{eq:finz}, we now get extra finite contributions to
$\hat Z$: 
\begin{align}
  \hat{Z'}^{(1,0)}_{ij,k} &= R_{im}R_{jn}\bigg(\hat{Z}^{(1,1)}_{mn,l} \tilde W_{ll'}
  \tilde U_{l'p} - \hat{Z}^{(1,1)}_{mn,l} \tilde U_{lp} + W_{ml}
  \hat{Z}^{(1,0)}_{ln,p} + W_{nl} \hat{Z}^{(1,0)}_{ml,p} \notag\\
  &\hspace{2.5cm} - W_{il} \hat{Z}^{(1,1)}_{ln,m} \tilde U_{mp} - W_{nl}
  \hat{Z}^{(1,1)}_{il,m} \tilde U_{mp} \bigg) \tilde R_{pk}^{-1} \, .
  \label{eq:fint1}
\end{align}
Here the indices $i$, $j$, and $k$ correspond to physical operators
only. These expressions have never been given explicitly in the
literature before. The anomalous dimension tensor then transforms
according to
\begin{align}\label{eq:chgt1}
  \gamma'^{(0)}_{ij,k} &= R_{im}R_{jn}\gamma_{mn,l}^{(0)} \tilde
  R_{lk}^{-1} \, , 
\end{align}
\begin{align}\label{eq:chgt2}
  \gamma'^{(1)}_{ij,k} &= R_{im}R_{jn}\gamma_{mn,l}^{(1)} \tilde
  R_{lk}^{-1} + \hat{Z'}_{ij,l}^{(1,0)}\tilde{\gamma'}_{lk}^{(0)} +
  2\hat{Z'}_{ij,k}^{(1,0)}\beta_0 +
  \hat{\gamma'}_{ij,l}^{(0)}\tilde{Z'}_{lk}^{(1,0)} \notag \\
  &\quad - \left\{ {\gamma'}_{il}^{(0)}\delta_{jm} +
    \delta_{il}{\gamma'}_{jm}^{(0)} \right\}\hat{Z'}_{lm,k}^{(1,0)} -
  \left\{ {Z'}_{il}^{(1,0)}\delta_{jm} + \delta_{il}{Z'}_{jm}^{(1,0)}
  \right\}\hat{\gamma'}_{lm,k}^{(0)} \, ,
\end{align}
as can be derived easily from Equations~\eqref{eq:chgtf1}
and~\eqref{eq:chgtf2}. A special case of these formulas has been
derived in Reference~\cite{Herrlich:1994kh}. Using the
definition~\eqref{eq:tw8}, we see that the dimension-eight Wilson
coefficients transform as
\begin{align}\label{eq:c8prime}
  \tilde C'_{k}(\mu) &= \tilde C_{i}(\mu) \tilde R_{ij}^{-1} \left [
    \delta_{jk} + \frac{\alpha_s (\mu)}{4 \pi} \tilde Z'^{(1,0)}_{jk}
  \right ] 
  \notag\\
  &\quad - C_{i}(\mu)R_{im}^{-1} C_{j}(\mu)R_{jn}^{-1} \bigg[ 
  \frac{\alpha_s (\mu)}{4 \pi} \hat Z'^{(1,0)}_{mn,k} \bigg] \, .
\end{align}

\subsection*{Transformation to the Traditional Operator Basis}

The calculation of the NLO QCD corrections to $\eta_{ct}$
in~\cite{Herrlich:1996vf} has been performed in a different basis for
the physical operators than the one chosen by us. It is given by
\begin{align}
{Q'}_1^{qq'} &= (\bar{s}_L^{\alpha}\gamma_{\mu}q_L^{\alpha})   
  \otimes (\bar{q'}_L^{\beta}\gamma^{\mu}d_L^{\beta})\, ,\notag\\
{Q'}_2^{qq'} &= (\bar{s}_L^{\alpha}\gamma_{\mu}q_L^{\beta})    
  \otimes (\bar{q'}_L^{\beta}\gamma^{\mu}d_L^{\alpha})\, ,\notag\\
{Q'}_3 &= (\bar{s}_L^{\alpha}\gamma_{\mu}d_L^{\alpha})   
  \otimes \sum\nolimits_q(\bar{q'}_L^{\beta}\gamma^{\mu}q_L^{\beta})\, ,\notag\\
{Q'}_4 &= (\bar{s}_L^{\alpha}\gamma_{\mu}d_L^{\beta})   
  \otimes \sum\nolimits_q(\bar{q'}_L^{\beta}\gamma^{\mu}q_L^{\alpha})\, ,\notag\\
{Q'}_5 &= (\bar{s}_L^{\alpha}\gamma_{\mu}d_L^{\alpha})   
  \otimes \sum\nolimits_q(\bar{q'}_R^{\beta}\gamma^{\mu}q_R^{\beta})\, ,\notag\\
{Q'}_6 &= (\bar{s}_L^{\alpha}\gamma_{\mu}d_L^{\beta})   
  \otimes \sum\nolimits_q(\bar{q'}_R^{\beta}\gamma^{\mu}q_R^{\alpha})\, . 
\end{align}
Note that we have expressed the operators in terms of left- and
right-handed fermion fields, in contrast to the definition used
in~\cite{Herrlich:1996vf}. The evanescent operators chosen
in~\cite{Herrlich:1996vf} are equivalent to the following set of
operators:
\begin{align}
{E'}_1^{qq'(1)} &= (\bar{s}_L^{\alpha}\gamma_{\mu_1\mu_2\mu_3}q_L^{\alpha})   
  \otimes (\bar{q'}_L^{\beta}\gamma^{\mu_1\mu_2\mu_3}d_L^{\beta}) - (16
  - 4 \epsilon){Q'}_1^{qq'} \, ,\notag\\
{E'}_2^{qq'(1)} &= (\bar{s}_L^{\alpha}\gamma_{\mu_1\mu_2\mu_3}q_L^{\beta})    
  \otimes (\bar{q'}_L^{\beta}\gamma^{\mu_1\mu_2\mu_3}d_L^{\alpha}) - (16
  - 4 \epsilon){Q'}_2^{qq'} \, ,\notag\\
{E'}_3^{(1)} &= (\bar{s}_L^{\alpha}\gamma_{\mu_1\mu_2\mu_3}d_L^{\alpha})   
  \otimes \sum\nolimits_q(\bar{q'}_L^{\beta}\gamma^{\mu_1\mu_2\mu_3}q_L^{\beta}) - (16
  - 4 \epsilon){Q'}_3 \, ,\notag\\
{E'}_4^{(1)} &= (\bar{s}_L^{\alpha}\gamma_{\mu_1\mu_2\mu_3}d_L^{\beta})   
  \otimes \sum\nolimits_q(\bar{q'}_L^{\beta}\gamma^{\mu_1\mu_2\mu_3}q_L^{\alpha}) - (16
  - 4 \epsilon){Q'}_4 \, ,\notag\\
{E'}_5^{(1)} &= (\bar{s}_L^{\alpha}\gamma_{\mu_1\mu_2\mu_3}d_L^{\alpha})   
  \otimes \sum\nolimits_q(\bar{q'}_R^{\beta}\gamma^{\mu_1\mu_2\mu_3}q_R^{\beta}) - (4
  + 4 \epsilon){Q'}_5 \, ,\notag\\
{E'}_6^{(1)} &= (\bar{s}_L^{\alpha}\gamma_{\mu_1\mu_2\mu_3}d_L^{\beta})   
  \otimes \sum\nolimits_q(\bar{q'}_R^{\beta}\gamma^{\mu_1\mu_2\mu_3}q_R^{\alpha}) - (4
  + 4 \epsilon){Q'}_6 \, . 
\end{align}
It turns out that in order to transform from our operator basis to the
traditional one the following four evanescent operators must be
introduced at the one-loop level in addition to the evanescent operators
given in Equation~\eqref{eq:Eoneloop} (see
Reference~\cite{Gorbahn:2004my}):
\begin{align}
E_5^{(1)}&=(\overline{s}_L\gamma_{\mu}
d_L)\otimes\sum\nolimits_q(\overline{q}\gamma^{\mu}\gamma_5 q) -
\frac{5}{3}Q_3 + \frac{1}{6}Q_5\, ,\notag\\
E_6^{(1)}&=(\overline{s}_L\gamma_{\mu}T^a
d_L)\otimes\sum\nolimits_q(\overline{q}\gamma^{\mu}\gamma_5 T^aq) -
\frac{5}{3}Q_4 + \frac{1}{6}Q_6\, ,\notag\\
E_7^{(1)}&=(\overline{s}_L\gamma_{\mu_1\mu_2\mu_3}
d_L)\otimes\sum\nolimits_q(\overline{q}\gamma^{\mu_1\mu_2\mu_3}\gamma_5 q) -
\frac{32}{3}Q_3 + \frac{5}{3}Q_5\, ,\notag\\
E_8^{(1)}&=(\overline{s}_L\gamma_{\mu_1\mu_2\mu_3}T^a
d_L)\otimes\sum\nolimits_q(\overline{q}\gamma^{\mu_1\mu_2\mu_3}\gamma_5 T^aq) -
\frac{32}{3}Q_4 + \frac{5}{3}Q_6\, ,
\end{align}

The transformation matrices $R$, $M$, $W$, and $U$ representing the
basis transformation according to Equation~\eqref{eq:gentmat}, as well
as the finite renormalisation induced by this transformation, can be
found in~\cite{Gorbahn:2004my}. The parts of the transformation matrices
relevant to us are given by\footnote{An additional rotation must be
  performed in order to change to the ``diagonal'' operator
  basis. This does not affect the finite renormalisation. }
\begin{equation}
R = 
\begin{pmatrix}
2 & \frac{1}{3} & 0 & 0 & 0 & 0 \\
 0 & 1 & 0 & 0 & 0 & 0 \\
 0 & 0 & -\frac{1}{3} & 0 & \frac{1}{12} & 0 \\
 0 & 0 & -\frac{1}{9} & -\frac{2}{3} & \frac{1}{36} & \frac{1}{6} \\
 0 & 0 & \frac{4}{3} & 0 & -\frac{1}{12} & 0 \\
 0 & 0 & \frac{4}{9} & \frac{8}{3} & -\frac{1}{36} & -\frac{1}{6}
\end{pmatrix}\, ,
\quad
M = 
\begin{pmatrix}
2 & \frac{1}{3} & 0 & 0 & 0 & 0 & 0 & 0 \\
 0 & 1 & 0 & 0 & 0 & 0 & 0 & 0 \\
 0 & 0 & 0 & 0 & 8 & 0 & -\frac{1}{2} & 0 \\
 0 & 0 & 0 & 0 & \frac{8}{3} & 16 & -\frac{1}{6} & -1 \\
 0 & 0 & 0 & 0 & -2 & 0 & \frac{1}{2} & 0 \\
 0 & 0 & 0 & 0 & -\frac{2}{3} & -4 & \frac{1}{6} & 1 
\end{pmatrix}\, ,
\end{equation}

\begin{equation}
W = 
\begin{pmatrix}
0 & 0 & 0 & 0 & 0 & 0 & 0 & 0 \\
 0 & 0 & 0 & 0 & 0 & 0 & 0 & 0 \\
 0 & 0 & 0 & 0 & 0 & 0 & 0 & 0 \\
 0 & 0 & 0 & 0 & 0 & 0 & 0 & 0 \\
 0 & 0 & 0 & 0 & -6 & 0 & 0 & 0 \\
 0 & 0 & 0 & 0 & 0 & -6 & 0 & 0
\end{pmatrix}\, ,
\quad
U =
\begin{pmatrix}
 0 & 0 & 0 & 0 & 0 & 0 \\
 0 & 0 & 0 & 0 & 0 & 0 \\
 0 & 0 & -112 & 0 & 16 & 0 \\
 0 & 0 & 0 & -112 & 0 & 16 \\
 0 & 0 & -\frac{10}{9} & 0 & \frac{1}{9} & 0 \\
 0 & 0 & 0 & -\frac{10}{9} & 0 & \frac{1}{9} \\
 0 & 0 & -\frac{136}{9} & 0 & \frac{10}{9} & 0 \\
 0 & 0 & 0 & -\frac{136}{9} & 0 & \frac{10}{9}
\end{pmatrix}\, ,
\end{equation}
whereas the matrix $V$ vanishes. They correspond to the bases 
\begin{equation}
Q' = ({Q'}_1^{qq'},{Q'}_2^{qq'},Q'_3,\ldots,Q'_6)\, , \quad
E' = \big({E'}_1^{qq'(1)},{E'}_2^{qq'(1)},{E'}^{(1)}_3,\ldots,{E'}^{(1)}_6\big)\, ,
\end{equation}
and
\begin{equation}
Q = ({Q}_1^{qq'},{Q}_2^{qq'},Q_3,\ldots,Q_6)\, , \quad
E = \big({E}_1^{qq'(1)},{E}_2^{qq'(1)},{E}^{(1)}_3,\ldots,{E}^{(1)}_8\big)
\end{equation}
in the notation of~\eqref{eq:gentmat}. The one-loop contribution to the
finite renormalisation in the dimension-six sector is given by
\begin{equation}
{Z'}_{QQ}^{(1,0)} = 
\begin{pmatrix}
 0 & 0 & 0 & 0 & 0 & 0 \\
 0 & 0 & 0 & 0 & 0 & 0 \\
 0 & 0 & \frac{178}{27} & -\frac{34}{9} & -\frac{164}{27} & \frac{20}{9} \\[1mm]
 0 & 0 & 1-\frac{{f}}{9} & \frac{{f}}{3}-\frac{25}{3} & -\frac{{f}}{9}-2 &
   \frac{{f}}{3}+6 \\[1mm]
 0 & 0 & -\frac{160}{27} & \frac{16}{9} & \frac{146}{27} & -\frac{2}{9} \\[1mm]
 0 & 0 & \frac{{f}}{9}-2 & 6-\frac{{f}}{3} & \frac{{f}}{9}+3 &
   -\frac{{f}}{3}-\frac{11}{3}
\end{pmatrix}\, .
\end{equation}
The finite renormalisation relevant for the mixing of dimension-six into
dimension-eight operators has never been calculated before. We find
\begin{equation}
\hat{Z'}_{QQ,\tilde Q_7}^{(1,0),T} = 
\begin{pmatrix}
0 & 0 & -20 & -\frac{20}{3} & 20 & \frac{20}{3} \\[1mm]
 0 & 0 & -\frac{20}{3} & -\frac{20}{3} & \frac{20}{3} & \frac{20}{3} 
\end{pmatrix}\, . 
\end{equation}

\subsection*{Transformation to the Diagonal Operator Basis}

Here we describe the change from the operator basis, where the
current-current operators are defined as in
Reference~\cite{Chetyrkin:1997gb, Gorbahn:2004my}, to
the diagonal basis, as defined in~\cite{Buras:2006gb} (and in this
work). The transformation matrices $R$, $M$, $U$, and $V$ in the
notation of~\eqref{eq:gentmat} are now given
by~\cite{Buras:2006gb,Gorbahn:2004my}\footnote{Here we have implicitly
  corrected some typos in Ref.~\cite{Buras:2006gb}. }
\begin{equation}\begin{split}
R = 
\begin{pmatrix}
1 & \frac{2}{3} & 0 & 0 & 0 & 0 \\[1mm]
 -1 & \frac{1}{3} & 0 & 0 & 0 & 0 \\[1mm]
 0 & 0 & 1 & 0 & 0 & 0 \\
 0 & 0 & 0 & 1 & 0 & 0 \\
 0 & 0 & 0 & 0 & 1 & 0 \\
 0 & 0 & 0 & 0 & 0 & 1
\end{pmatrix}\, ,
&\quad
M_{ij} = 
\begin{cases}
1, &i=j\, ,\\
20, &(i,j) \in \{(9,1),(10,2)\}\, ,\\
0, &\text{otherwise};
\end{cases}
\\[3mm]
U_{ij} =
\begin{cases}
4, &(i,j) \in \{(1,1),(2,2)\}\, ,\\
144, &(i,j) \in \{(5,1),(6,2)\}\, ,\\
0, &\text{otherwise};
\end{cases}
&\quad
V_{ij} =
\begin{cases}
4, &(i,j) \in \{(1,1),(2,2)\}\, ,\\
\frac{3712}{25}, &(i,j) = (5,1)\, ,\\
\frac{8032}{25}, &(i,j) = (6,2)\, ,\\
0, &\text{otherwise};
\end{cases}
\end{split}\end{equation}
and the matrix $W$ vanishes. These matrices correspond to the following
bases of operators (the roles of the primed and unprimed set of
operators is reversed with respect to Reference~\cite{Buras:2006gb}):
\begin{equation}
Q' = (Q_+,Q_-)\, , \quad
E' = \big(E^{qq'}_1,E^{qq'}_2,E^{qq'}_3,E^{qq'}_4\big)\, ,
\end{equation}
and 
\begin{equation}
Q = (Q_1, Q_2) \, , \quad
E = \big(E_1^{(1)},E_2^{(1)},E_1^{(2)},E_2^{(2)}\big)\, .
\end{equation}
All necessary renormalisation constants can be found in
Reference~\cite{Gorbahn:2004my}. The finite renormalisation is then
given by
\begin{equation}
{Z'}_{QQ}^{(1,0)} = 
\begin{pmatrix}
-\frac{5}{3} & -\frac{8}{9} & 0 & 0 & 0 & 0 \\
 -4 & 0 & 0 & 0 & 0 & 0 \\
 0 & 0 & 0 & 0 & 0 & 0 \\
 0 & 0 & 0 & 0 & 0 & 0 \\
 0 & 0 & 0 & 0 & 0 & 0 \\
 0 & 0 & 0 & 0 & 0 & 0
\end{pmatrix}\, ,
\quad
{Z'}_{QQ}^{(2,0)} = 
\begin{pmatrix}
-\frac{29123}{900}-\frac{25}{54}f &
   \frac{17}{135}-\frac{20}{81}f & 0 & \frac{11}{27} & 0 &
   0 \\[2mm]
 -\frac{343}{30}-\frac{10}{9}f & -\frac{498}{25} & 0 &
   -\frac{4}{9} & 0 & 0 \\
 0 & 0 & 0 & 0 & 0 & 0 \\
 0 & 0 & 0 & 0 & 0 & 0 \\
 0 & 0 & 0 & 0 & 0 & 0 \\
 0 & 0 & 0 & 0 & 0 & 0
\end{pmatrix}\, .
\end{equation}



\end{document}